\documentclass[]{spie}  
\usepackage{amsmath,amsfonts,amssymb}
\usepackage{graphicx}
\usepackage{setspace}
\usepackage{tocloft}
\usepackage[table]{xcolor}
\usepackage{todonotes}
\usepackage{caption}
\usepackage{subcaption}
\usepackage{wrapfig}
\usepackage{tcolorbox}

\title{Sub-percent Characterization and Polarimetric Performance Analysis of Commercial Micro-polarizer Array Detectors}

\author[a,*]{Thijs Stockmans}
\author[a]{Naor Scheinowitz}
\author[b]{Erwoud van der Linden}
\author[b]{Irina Malysheva}
\author[a]{Kira Strelow}
\author[b]{Martijn Smit}
\author[a]{Frans Snik}

\affil[a]{Leiden Observatory, Leiden University, P.O. Box 9513, 2300 RA Leiden,
The Netherlands}
\affil[b]{SRON Netherlands Institute for Space Research, Niels Bohrweg 4,
2333 CA Leiden, The Netherlands}

\cftpagenumbersoff{figure}
\cftpagenumbersoff{table} 
\begin{document} 
\maketitle
\pagenumbering{arabic}

\begin{abstract}
Polarization imaging can yield crucial information in multiple applications of remote sensing, such as characterization of clouds, aerosols, and the Aurora Borealis. 
Some applications require sub-percent polarimetric sensitivity and accuracy in determining the Stokes parameters, which can be a challenge to attain. 
In 2018, Sony released a low-cost CMOS-based imaging chip with integrated micro-polarizer array for general polarization measurements. 
We implement the calibration steps required for these Sony chips to reach sub-percent polarimetric accuracies. 
To analyze their performances, we have compared the characteristics of four different detector packages by three manufacturers housing either the monochromatic version or the RGB color variant. 
We present a comprehensive overview of the effects that these characteristics have on the polarimetric performance of the camera. They include dark noise, behavior over different gain settings, detector/pixel artifacts, and polarimetric effects determined by polarizer extinction ratios, polarizer orientations, and accuracy of polarimetric zero points due to differential pixel gains. 
In addition to calibrations using unpolarized light and fully linearly polarized light, we assess the polarimetric sensitivity within a tilting and rotating glass-plate set-up. 
We discuss the benefits of adding a rotating half-wave plate as an additional temporal modulator to generically mitigate some of the detector effects, and achieve better polarimetric sensitivity/accuracy albeit at the expense of lower temporal resolution. 
We conclude by presenting and discussing the polarimetric limits to which we were able to calibrate the detector effects for practical purposes. 
By reaching a compound absolute polarimetric uncertainty of less than a percent, these very compact, low-cost detectors are enabled for a multitude of scientific goals.
\end{abstract}


\section{INTRODUCTION}

Imaging polarimetry in the optical regime (UV -- mid-IR) is an invaluable and still developing technique for remote detection and characterization of a range of sources, in research fields ranging from astronomy to Earth observation, and from biomedical imaging to target detection \cite{tyo_review_2006, chenault_overview_2014}.
Particularly when combined with wavelength-selectivity (e.g.~spectrally resolved or multispectral observation) and time resolution, all relevant aspects of such sources can be acquired and consequently analyzed \cite{rodenhuis_five-dimensional_2014}.
As common detectors are generally insensitive to the polarization state of the incident light, the implementation of polarimetric functionality\cite{snik_astronomical_2013} was often considered cumbersome.
The most basic solution consists of a rotating polarizer in front of an imaging detector, but this renders the polarization measurement susceptible to any variation or relative movement of the source.
The use of one or several polarizing beam-splitters can enable instantaneous polarization measurements, but requires complex optical paths and the use of divided detector space or even several detectors.
Moreover, this approach is limited by any differences between the split beams (transmissions, pixel gains, alignments, aberrations).
The systematic effects that limit both these implementations can be mitigated by at least an order of magnitude by combining the spatial splitting with temporal modulation in the form of a rotating half-wave plate in front of a single polarizing beam-splitter.
This implementation does not only allow for intensity filtering conform the complete linear Stokes parameters, but its ``dual-beam exchange'' also enables a redundant measurement scheme involving beam-switching.
A double-difference demodulation then yields the fractional linear Stokes parameters for which all systematic effects between the beams and in time divide out\cite{bagnulo_stellar_2009, snik_astronomical_2013}.

A compact and robust technical implementation for polarimetric imaging was introduced by aligning and bonding polarizer filters on a per-pixel level. 
These sensors have acquired multiple names in the literature, amongst which some examples are: ``division of focal plane'' (DoFP) polarimeter, ``microgrid'' polarimeter, or a ``micropolarizer array'' (MPA) detector.
The first to develop and patent this concept was D.M.~Rust in 1995. \cite{rust_integrated_1995}
One of the first publications to describe the fabrication of a pixelated polarizer filter, and successively create a micropolarizer array detector were Kalayjian et al.~in 1996. \cite{kalayjian_polarization_1996}
Since 2D techniques for this polarization pattern did not exist yet at the time, they kept to 1D strips of chemically etched polarizing films on top of the photodiode array, which were only sensitive to the first two Stokes parameters.
The initial idea and fabrication of a micropolarizer array was improved upon by multiple groups afterwards. \cite{guo_fabrication_2000, maruyama_32-mp_2018}
Other polarization filters were introduced, like birefringent crystals, photolithographied films and aluminium wiregrids. \cite{momeni_analog_2006, zhao_thin_2009, gruev_ccd_2010}
The additional polarization filters at 45 an 135 degrees were also introduced, making the sensor sensitive to the full first 3 Stokes parameters. \cite{gruev_image_2008}.  
Similar detectors were developed in the mid-IR. \cite{chenault_infrared_2016}.
Advancements in metamaterials is deeply connected due to the micro-fabrication of these detectors, pushing the concept to other wavelength regimes, from the deep ultraviolet to THz radiation. \cite{zou_ultra-broadband_2020, zhang_deep-ultraviolet_2019}
By combining them with patterned liquid-crystal polymers, these detectors were upgraded to provide full-Stokes measurements.\cite{hsu_full-stokes_2014, tu_division_2020}

The development of these micropolarizer arrays has continued to this day. \cite{guan_integrated_2022}  
When the original patent expired in 2013, the MPA detectors first became commercially available in the visible-light regime from 4D.\cite{brock_snap-shot_2014, vorobiev_astronomical_2018}
The 4D sensor was used for important applications in interferometry.\cite{millerd_vibration_2017}
In 2018, Sony released its first CMOS-based polarization imaging sensors, which integrated patterned wiregrid polarizers on-chip onto the pixels in the lithography process.\cite{maruyama_32-mp_2018}
Nowadays, the Sony IMX250MZR/MYR, IMX253MZR/MYR and IMX264MZR/MYR sensors are commonly available in various electronics packages from several vendors, with options for monochrome and RGB color, and a range of detector formats and read-out speeds.
The availability of these sensors as a component-off-the-shelf (COTS) has sprouted an increase in academic interest in these DoFP sensors, as they can be easily incorporated into an optical set-up to replace polarization-insensitive imaging detectors. 
Because of its size, the detector can be easily connected with an unmanned aerial vehicle to perform, for instance, evidence searches and solar-farm inspections.\cite{sun_unmanned_2024, takacs_polarized_2024}
The sensor was used to remove water glints in surface water imaging. \cite{wang_automatic_2023}
By having access to 2 other information dimensions on a detector level, videos can be made of ultrashort laser pulses. \cite{inoue_motion-picture_2022}
By combining the detector with notch filters and computational imaging techniques, multi spectral-polarimetric imaging was enabled. \cite{huang_high-efficiency_2023}
Combining with a double dispersive element and a digital micro-mirror device, compressive sensing also enables spectropolarimetric imaging. \cite{zhang_dual-dispersive_2024}

While these sensors are extremely useful for easy implementation of polarimetric imaging at moderate polarimetric accuracy, two important challenges remain:
\begin{enumerate}
\item Because the polarimetric demodulation is constrained by the intensity measurements of four separate polarization-filtered pixels, intensity structures across the field-of-view can induce spurious polarization signals through aliasing.\cite{tyo_total_2009}
\item Because of prevalent differential pixel gains and other detector effects and the imperfections of the micropolarizers' orientations and extinction ratios, the intrinsic polarimetric accuracy of the detectors is limited, both in terms of the zero point of the fractional Stokes parameters, as well as in terms of their polarization response.\cite{snik_astronomical_2013, chenault_overview_2014}
\end{enumerate}

The fundamental issue of aliasing (also known as the ``IFOV effect'') can be mitigated by applying sufficient defocus or other manipulation of the Modulation Transfer Function, which loses spatial resolution. \cite{ratliff_interpolation_2009}
However, in-focus images are widely used and the missing spatial information is interpolated between pixels.\cite{ratliff_image_2006}
The most basic form of demodulating the polarization information from the intensity measurements is through linearly combining the intensities at a ``superpixel'' level.
Where a superpixel is defined as a recurrent block of 2x2 pixels that contains all four different polarization-filtering angles.
However, the retrieval of the polarization information has improved beyond that and has been closely linked to the field of ``demosaicing'' of regular Bayer-pattern RGB images, or ``super-resolution''.\cite{hardie_super-resolution_2011, mihoubi_survey_2018}
Several authors have provided different solutions and algorithms to conduct the demosaicing of MPA sensors. 
A more comprehensive overview can be found in the review paper from Mihoubi et al.~(2018)\cite{mihoubi_survey_2018}, but we present some examples below.
Interpolation in the image plane can be done using bilinear interpolation or cubic interpolation or a combination of them considering local smoothness. \cite{ratliff_image_2006, ratliff_interpolation_2009, zhang_novel_2017}
Others used Gaussian processes to do the interpolation and combine it with an additional denoising.\cite{gilboa_image_2014}
The denoising algorithms were, in turn, expanded upon to include non-gaussian noise sources.  \cite{wang_image_2018}, and an overall analysis of the impact of noise on reconstruction/demosaicing algorithms was performed.\cite{bai_noise_2022}
Making use of the correlation between superpixels, others have used guided filtering to remove aliasing effects. \cite{liu_new_2020}
The newest forms of interpolation make use of residual interpolation with a range of flavors: edge-aware, intensity-guided, Modified Newton, or iterative. \cite{morimatsu_monochrome_2020, morimatsu_monochrome_2021, liu_modified_2022, yang_residual_2022} 

In addition, the standard ``superpixel'' demodulation can be replaced by Fourier methods to separate the bands carrying intensity and polarization information\cite{tyo_total_2009, hagen_fourier-domain_2024}.
Moreover, commercially available detectors apply 2$\times$2 superpixelation, analogous to RGB sensors, but it has been demonstrated that a 2$\times$4 pattern is more favorable to use the space in the Fourier domain.\cite{lemaster_improved_2014}
Over the years, even more intricate patterns have been developed to increase the bandwidth and improve accuracy. \cite{alenin_optimal_2017, hao_new_2021, ratliff_alternative_2021}
The RGB versions of the Sony detector are manufactured using a (2$\times$2)$\times$(2$\times$2) repeating superpixel pattern of 4$\times$4, while also for such configurations more optimal combined color/polarization solutions can be found\cite{vaughn_spatio-temporal_2019}.

In this work, we have focused our efforts on mitigating the detector effects that intrinsically limit the polarimetric accuracy at fractional polarization levels of $\sim$1\% (or worse), and, through rigorous calibration, achieve a sub-percent polarimetric accuracy.
Most importantly, we aim to obtain a fundamental understanding of the limiting effects, and their mutual interactions.
Moreover, we demonstrate that additional modes of polarization modulation (e.g.~in the temporal domain with a rotating half-wave plate\cite{snik_multi-domain_2015, vaughn_portable_2015}) can generically compensate for systematic effects present in such micropatterned polarization cameras.

We are not the first to conduct a calibration and characterization of these types of detectors in general or of the commercially available ones. \cite{gimenez-henriquez_characterization_2022}
The first description of three set-ups for the characterization of these MPA detectors was given by York and Gruev (2012), two of which have been the base of the characterization measurements that we conduct in this paper. \cite{york_characterization_2012}
They describe a set-up containing a integrating sphere to measure unpolarized light, with uniform illumination.
The third set-up mentioned in the paper combines a wiregrid polarizer and quarter waveplate to shine known fully polarized light on the detector. 
They describe how to measure the accuracy in measuring the degree and angle of linear polarization (DoLP, AoLP) and the spectral response using these set-ups in a pixel-to-pixel basis. 
The main calibration scheme for the DoFP cameras was described by Powell and Gruev (2013).\cite{powell_calibration_2013}. 
The superpixel calibration scheme described in this paper can correct for photodetector gain differences, and mitigate non-ideal properties like filter and diattenuation coefficients / extinction ratios. 
Multiple variations of this calibration scheme have been proposed in the years that followed. 
Gimenez et al.~(2020) bundled the available calibration schemes and tested them relative to each other. \cite{gimenez_calibration_2019, gimenez_calibration_2020}
Their conclusion was that the superpixel technique described by Powell and Gruev still had the best performance. 
Originally this calibration is done for each individual superpixel, but it has been shown that with relatively small error, the calibration can be done with a general correction matrix.\cite{lane_calibration_2022}
A simplified scheme of calibration, describing it fully in terms of diattenuation, misalignment and estimated light intensity, was provided by Hagen et al.~(2019). \cite{hagen_calibration_2019}
Their measurement set-up consisted of fully linear polarization at 4 angles.
They mainly showed that the measurement of the extinction ratio is very prone to error. 
Luckily, even with bad extinction ratios, it is still possible to perform good measurements of the Stokes parameters as long as a good calibration accuracy is achieved, and the measured degree of linear polarization is relatively low. \cite{hagen_generating_2019}
In an even simpler calibration scheme, the incident polatization angle does not need to be known exactly, as long as it is rotated in steps of 90 degrees. \cite{rodriguez_practical_2022}

An important feature of these detectors is their spectral response. 
Several of the correction parameters described above vary with wavelength, across the ``monochrome'' wavelength range ($\sim$300--1000 nm) and the RGB filter bandpasses\cite{szaz_drone-based_2023}. 
The relative spectral responses can exhibit variation by an order of magnitude over the visible wavelength range, rendering several of the calibration parameters sensitive to the spectral features of the light sources and/or the target. \cite{venkatesulu_measuring_2022}

The micro-polarizer array detector, like all photodetectors, is fundamentally limited by noise, of which shot noise is the most prominent in many cases. 
The propagation of these noise sources to the polarization measurement has been described\cite{roussel_polarimetric_2018}, taking into account both Poissonian and additive noise, and verification with a Monte-Carlo code\cite{chen_analysis_2021}.

This paper introduces a comparison of four different COTS imaging detectors which all contain a DoFP polarization sensor from Sony. 
We perform the essential detector and polarization calibration steps, and present a more in-depth analysis of the effects of the dark noise and non-linearity and their calibration approaches, also with the specific aim of implementation in the field, and therefore under unstable (temperature) conditions. 
For each of the individual systematic/random effects of the detector and its micropolarizer array that we calibrate and analyze, we highlight the propagation of their residuals to the uncertainty (i.e.~absolute polarimetric sensitivity and/or accuracy) on the observable fractional linear Stokes parameters $[Q/I,U/I]$, for a typical practical case in terms of illumination, polarization signal and temperature-stability.
Where relevant, we compare the performance of the four different detector packages that we have calibrated.
Finally, we compare the polarimetric performance of a fully calibration MPA detector to a system with ``spatio-temporal modulation'' through an addition of a rotating half-wave plate. 
This analysis is performed for both practically unpolarized light from an integrating sphere and a controlled increase in partially polarized light using a tilted glass-plate set-up.  

With our calibration and mitigation strategies to reach down to $\sim$10$^{-3}$ polarimetric accuracy, we open up the Sony polarization sensor to more challenging and advanced science cases, which include:
\begin{itemize}
\item Atmospheric aerosol characterization through multi-angle and multi-wavelength polarimetric observations of scattered sunlight. Currently, only the SPEXone multi-angle spectropolarimeter\cite{snik_spectral_2009, smit_spex_2019, rietjens_spexone_2023} instrument onboard the PACE satellite achieves the $\sim$10$^{-3}$ polarimetric accuracy necessary to retrieve the full aerosol microphysical parameters, including complex refractive index\cite{hasekamp_retrieval_2007, dubovik_polarimetric_2019}. The PACE/HARP2 instrument\cite{mcbride_spatial_2019} implements full hyperangular imaging over a large field-of-view, at the cost of polarimetric accuracy. The micropatterned polarizer imaging detector may provide a compact solution for cubesat\cite{levis_3d_2021} and ground-based observation, provided sufficient calibration accuracy and stability.
\item Some of the auroral emission lines are likely linearly polarized due to the anisotropic excitation by precepitating electrons. However, this polarization is small ($\sim$1\% for the red line), and the auroral displays are usually faint and highly variable.\cite{lilensten_thermospheric_2013, barthelemy_measurement_2019, bosse_nightglow_2020} Hence, a snapshot polarimetric imaging approach using a well-calibrated pixelated polarizer array detector can be beneficial, particularly when the RGB pixels can be uniquely spectrally mapped onto the red, green and blue auroral lines due to atomic oxygen and molecular nitrogen, respectively.
\item Micropolarizer detectors can also be used to enable imagery of absorbsion due to a range of atmospheric trace gases. By implementing a birefringent filter, the average absorption strength across a molecular spectral band is converted into a fraction polarization signal, which is then readily recorded by a polarization camera\cite{derkink_imaging_2022}. However, the variations in concentration (or path length) of such trace gases are typically small, such that an absolute spectral/polarimetric sensitivity of $\sim$10$^{-3}$ is required for NO$_2$ sensing in the 425--450 nm spectral range, which is accessible to the Sony chips.
\end{itemize}

This paper follows the flow of the calibration scheme that we propose to push the MPA sensors to their best possible polarimetric accuracy. 
We discuss all relevant measured effects and propagate their individual impacts on the measurement of the Stokes parameters. 
We show some combined effects of all residual error terms, and compare that to the performance using additional temporal polarimetric modulation. 
Finally, we bundle our conclusions and discuss our recommendations in the corresponding final section.

\section{CALIBRATION SET-UP \& CAMPAIGN}

\subsection{Used Detectors and their Specifications}
We conduct our measurements using 4 different polarization cameras, where the electronics are created by three commercial companies: FLIR, LUCID, and Thorlabs.  
They are all fitted with a chip from the Sony series. 
We have listed the main specifications of interest in table \ref{tab: camera specs}. 

\begin{table}[]
\caption{Summary of the used cameras and their main specifications used for this paper}
\label{tab: camera specs}
\begin{tabular}{|l|c|c|c|c|}
\hline
                   & \begin{tabular}[c]{@{}c@{}}FLIR \\ BFS-U3-51S5PC-C\end{tabular}                     & \begin{tabular}[c]{@{}c@{}}LUCID \\ TRI050S1-QC\end{tabular} & \begin{tabular}[c]{@{}c@{}}FLIR \\ BFS-U3-51S5P-C\end{tabular}                      & \begin{tabular}[c]{@{}c@{}}Thorlabs\\ Kiralux CS505MUP\end{tabular} \\ \hline
Sony Chip          & IMX250MYR                                                                           & IMX264MYR                                                    & IMX250MZR                                                                           & IMX264MZR                                                           \\ \hline
monochrome/RGB     & RGB                                                                                 & RGB                                                          & Mono                                                                                & Mono                                                                \\ \hline
temperature sensor & yes                                                                                 & yes                                                          & yes                                                                                 & no                                                                  \\ \hline
max bitsize        & 12                                                                                  & 12                                                           & 12                                                                                  & 12                                                                  \\ \hline
max frame rate     & \begin{tabular}[c]{@{}c@{}}12 bit: 89.5 \\ 10 bit:144.7\\ 8 bit: 163.4\end{tabular} & 12 bit: 35.7                                                 & \begin{tabular}[c]{@{}c@{}}12 bit: 89.5 \\ 10 bit:144.7\\ 8 bit: 163.4\end{tabular} & 12 bit: 35.7                                                        \\ \hline
Frame size         & 2448x2048                                                                           & 2448x2048                                                    & 2448x2048                                                                           & 2448x2048                                                           \\ \hline
max analog gain    & 24                                                                                  & 24                                                           & 24                                                                                  & 24                                                                  \\ \hline
\end{tabular}
\end{table}


\subsection{Dark Offset Measurements}
To measure the dark offset as a function of temperature, we placed the detectors in a climate-controlled chamber. 
The climate-controlled chamber can regulate the temperature between 173 and 473 $^\circ$Kelvin with an accuracy of .1 $^\circ$K. 
The cameras were shielded from all light sources, so that no photons could reach the detectors and only the dark offset remained.
In order to build a valid model of the dark offset, we took measurements at multiple exposure times between the maximum and minimum frame rate, gain settings from 0 to 24 and various temperatures falling within the typical operation range (typically 0-40 degrees Celsius) of the cameras. 
The acquisition scheme of each camera was estimated to sample as much as possible in the diverse ranges within 90 minutes per fixed external temperature.
We made use of the internal temperature sensor if possible to link each dark offset measurement with a temperature of the camera.
The internal temperature fluctuates due to latency or heating up due to operations, so for each fixed external temperature, there is an internal temperature range as big as 5 degrees Kelvin.
These fluctuations resulted in over a hundred measurement sets per camera with slightly different settings or temperatures.
Since these setting combinations are unique to each camera, we cannot give a comprehensive overview of all the settings for each of these measurement sets.
Instead, we note in table \ref{tab:dark_measurements} the measurement scheme for the LUCID camera to give an example of the experiment.

\begin{table}[]
\caption{Dark offset measurement settings for the LUCID camera.}
\label{tab:dark_measurements}
\small
\begin{tabular}{|l|cccccccc|}
\hline
\textbf{Gains}                                                                        & 0          & 0.1                  & 0.5                  & 1                    & 3                    & 13                   & 23                   & 24                   \\ \hline
\textbf{\begin{tabular}[c]{@{}l@{}}External \\ temperatures\\ (Celcius)\end{tabular}} & -20        & -10                  & -5                   & 0                    & 10                   & 20                   & 30                   & 40                   \\ \hline
\textbf{Exposure times ($\mu$s)}                                                          & 80         & 100                  & 300                  & 500                  & 700                  & 1000                 & 3000                 & 5000                 \\ 
                                                                                      & 10 000     & 50 000               & 100 000              & 500 000              & 1 000 000            & 2 000 000            & 3 000 000            & 5 000 000            \\
                                                                                      & 10 000 000 \\ \hline
\end{tabular}
\end{table}

\subsection{Pixel Gain Calibration}
To correctly quantify the noise of the detector it is necessary to make an estimation of the electron to digital counts conversion factor as a function of the user defined gain. 
We have fitted a photon transfer curve of Digital Counts versus the standard deviation of the Digital Counts for constant unpolarized illumination, and varying integration times.
We find that the conversion factor g follows the following formula: 
\begin{equation}
    g = 0.36 \times 10^{\frac{G}{20}}
\end{equation}
This is in line with the measurements of Chen et al. 2021. \cite{chen_analysis_2021}

\subsection{Collimated Beam Set-up for Polarization Calibration} \label{sec:setup} 
For the polarimetric characterization of the cameras, a setup based on an integrating sphere illuminated with a Laser-Driven Light Source (LDLS) XWS-30 was used.
This set-up is similar to the the first set-up described in York and Gruev 2012. \cite{york_characterization_2012}
The 2" integrating sphere (Edmund Optics \#58-584) produces practically unpolarized light due to the multiple reflections of the internal spectralon coating. \cite{mcclain_depolarization_1995} 
The output light is collimated with a 10 cm achromatic lens (Thorlabs AC254-100-A-ML) to be able to place color filters in the beam. 
The central wavelength of the filters varied from 450 nm to 650 nm with steps of 100 nm, each with a FWHM of 40 nm. The central wavelengths were chosen to match the reported peak quantum efficiency for the respective red, green and blue pixels on the RGB sensitive cameras. 
Then, the second lens with $F = 20$ cm focused the beam on the diaphragm of 4 mm, located in the focal plane of the third lens with $F = 30$ cm. 
This composition resulted in a collimation of 1 degree of the output beam. The beam-size is a factor of $2.5$ bigger than the detector itself to mitigate the edge effects such as intensity drop-offs.  

\begin{figure}
    \centering
    \includegraphics[width=0.85\linewidth]{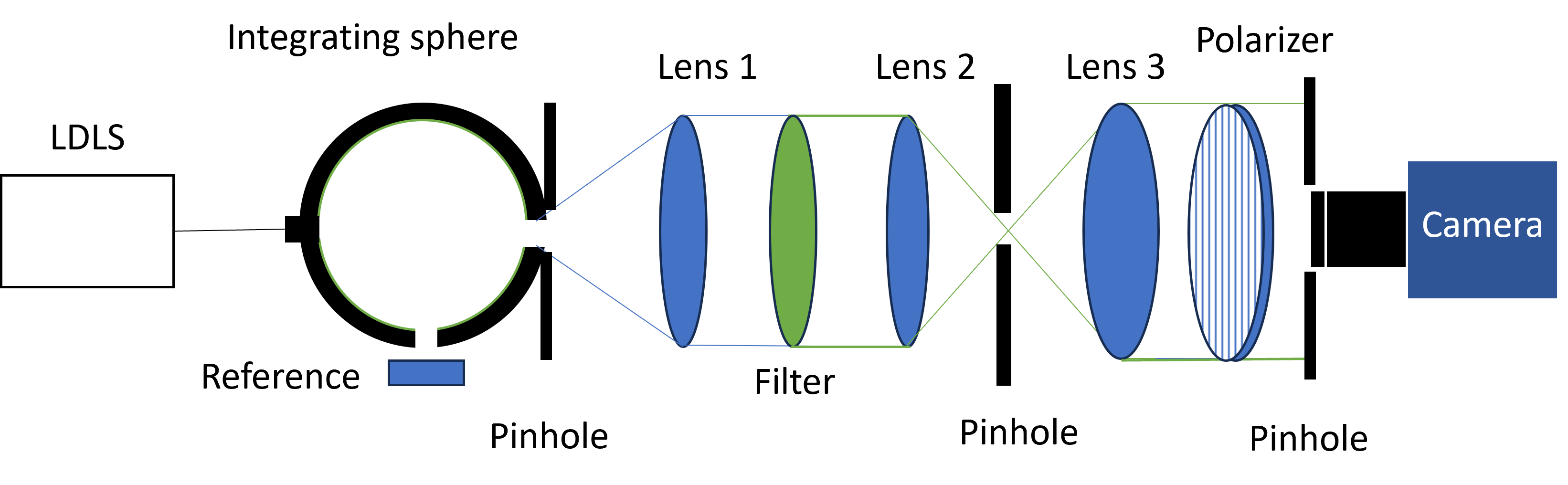}
    \caption{From integrating sphere to detector}
    \label{fig:collimated beam set-up}
\end{figure}

However, lenses add a polarization profile to the beam in a radial manner, leading to the light reaching the detector being slightly polarized. \cite{chipman_polarization_1986} In addition, stress birefringence in the lens glasses may modify this induced polarization.
To get an upper bound on the degree of linear polarization in the experimental setup, we measured a beam profile using one of the four polarization cameras.

There is some scientific tension in using the device that is up for calibration to verify the performance of the setup meant to calibrate it, so we will merely interpret the resulting value as an indication of the worst case scenario. 

\begin{figure}
    \centering
    \includegraphics[width=0.55\linewidth]{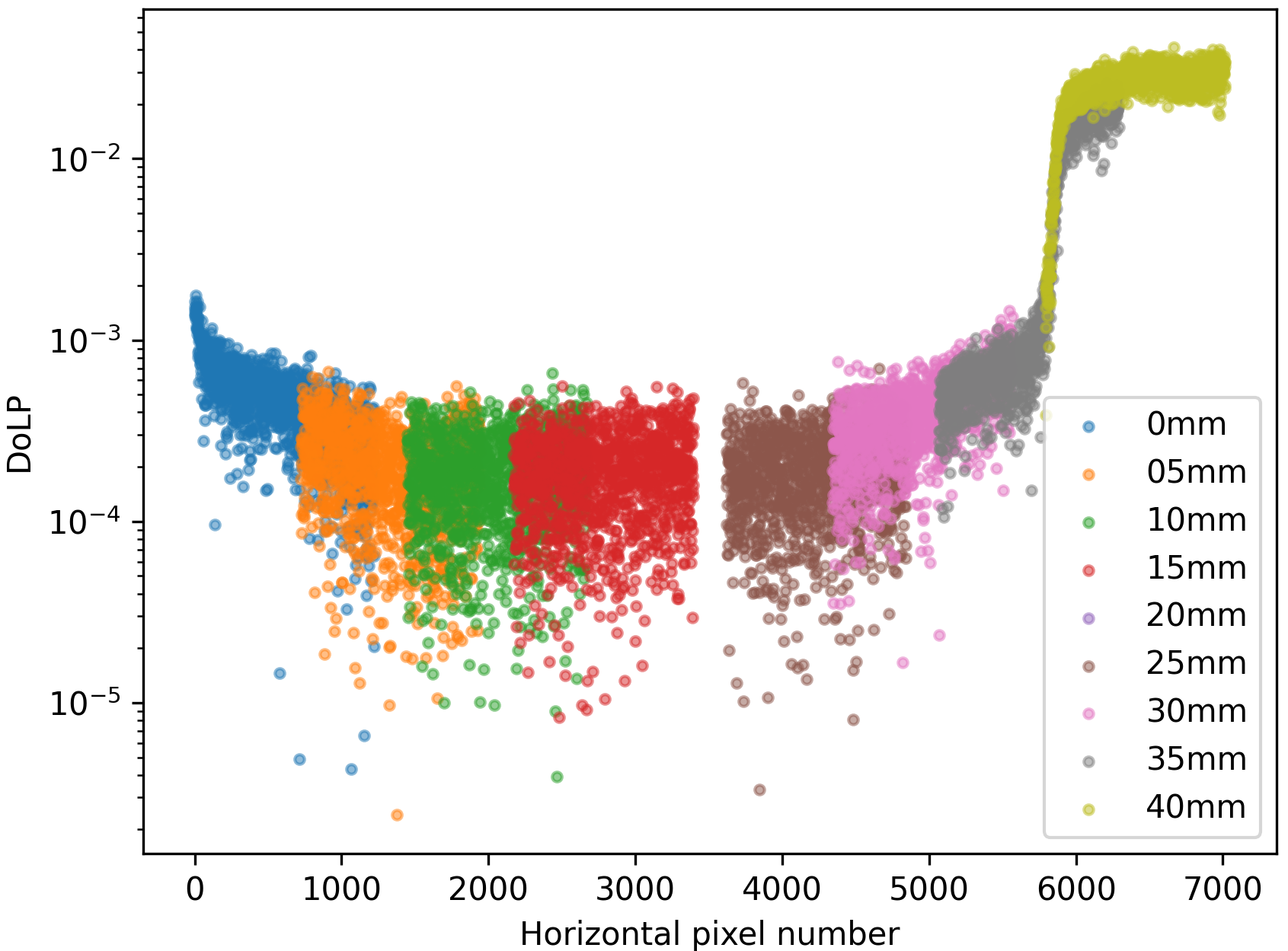}
    \caption{Polarization profile of the collimated beam comprised of 9 images along a horizontal sweep of the beam. The horizontal axis represents the column indices of the combined profile after combining the individual images. Each circle in the plot represents the column-averaged DoLP value.}
    \label{fig:col_beam_imperfection}
\end{figure}

We mounted the camera on a translation stage perpendicular to the optical axis and moved it steps of 5 mm from one edge of the beam horizontally to the other. 
As the sensor is larger than 5 mm we could overlap the data to create a full profile of the beam in terms of DolP, as shown in figure \ref{fig:col_beam_imperfection}. 
While analyzing the data, we noted an identical gradient in each individual image, so we used the center image (at 20mm) as a flat field for the other images to ensure any visible effects are due to the beam and not the CMOS sensor.
Based on the results we measured, we conclude that the remaining polarization in the collimated beam is at most $10^{-3}$ in the center of the beam. 

For the measurements that required fully linearly polarized light we added a linear polarization filters as the last optical element.
The polarizer of choice was a 2 inch Edmund Optics \#19-652 VIS coated, linear glass polarizer, based on XP42 polarizer film. 
According to the manufacturer's specifications, the polarizer performance is not homogeneous over the entire wavelength range we are interested in.
To mitigate this inhomogeneity, we used spectral filters in all our measurements with polarized light. 
We assume our spectral filters to be sufficiently narrow that any mixing or leaking effects are negligable.
Furtheremore, the manufacturer claims an extinction ratio of 9000:1, which is an order of magnitude larger than what the camera manufacturers claim for their extinction ratio. 
We therefore assume that we can disregard the calibration polarizer as a limiting factor in measuring the extinction ratio.

\subsection{Glass Plate Set-up}\label{sec: glass plate}
\begin{figure}
    \centering
    \includegraphics[width=0.85\linewidth]{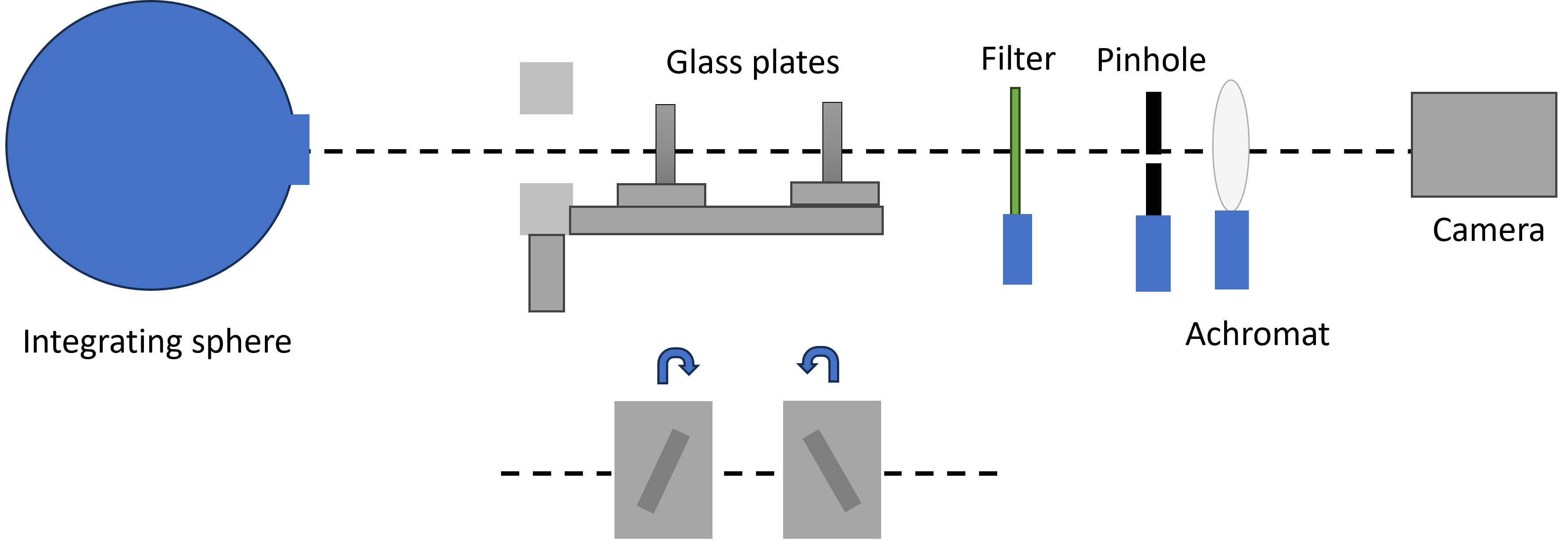}
    \caption{Glass plates set up}
    \label{fig:Glass_plates}
\end{figure}

\begin{figure}
    \centering
    \includegraphics[width=0.55\linewidth]{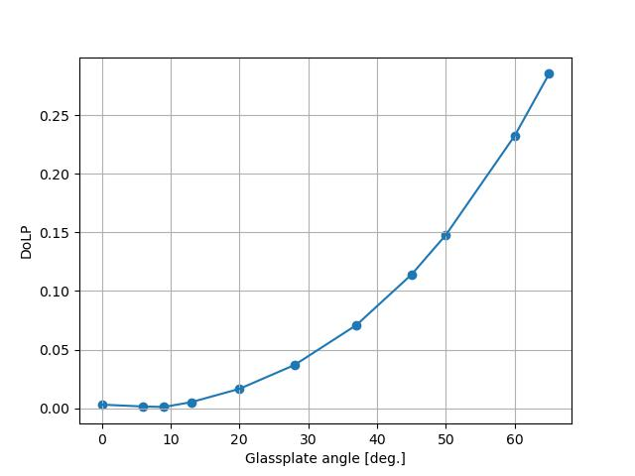}
    \caption{DoLP dependence on the angle of the glass plate.}
    \label{fig:dolp_angle_glass_plates}
\end{figure}

To measure partially polarized linear polarization, we made use of an existing set-up with two glass plate of well-known properties, mounted in motorized rotatable tilt mechanisms (fig.~\ref{fig:Glass_plates}).\cite{smit_spex_2019, harten_calibration_2018}
In this set-up, practically unpolarized light comes from the integrating sphere (the same as in the set-up in section \ref{sec:setup}). 
This light is directed through two glass plates which are tilted with respect to each other along the same (rotatable) axis. 
For the wavelength selection a color filter with central wavelength of 550 nm and FWHM of 40 nm is used (Thorlabs FBH550-40).  
The stray light was shielded with the pinhole with opening of $\approx5$ mm. 
The beam is collimated on the detector with an achromatic lens with $F = 7.5$ cm (Thorlabs AC508-075-A).
The magnitude of the tilt directly relates to the degree of linear polarization (fig. \ref{fig:dolp_angle_glass_plates}). 
This relation was verified with the use of CMOS camera (FLIR BFS-U3-17S7M-C) for the measurement and the broadband wire-grid polarizer (Thorlabs).
This measurement is used to correct the zero points in $Q/I$ and $U/I$ to sub-percent accuracy.

\section{POLARIMETRIC CALIBRATION AND ANALYSIS FOR INDIVIDUAL EFFECTS}

In this section we describe the polarimetric impact of all individual sensor effects that impact the polarimetric accuracy and/or the polarimetric sensitivity of the sensors. 
We will mostly describe the impact in terms of the measurement fractional Stokes parameters, since the mean of these parameters follow a Gaussian-like noise profile under ideal circumstances, compared to the skewed noise distributions that the Degree of Linear Polarization and Angle of Polarization follow. 
To define the mathematical formalism for linear polarized light in terms of the Stokes parameters: 

\begin{align}
    \begin{bmatrix}
I \\
Q \\
U \\
V 
\end{bmatrix}  
&=
\begin{bmatrix}
\frac{1}{4}(I_0+I_{90}+I_{45}+I_{135}) \\
I_0-I_{90} \\
I_{45}-I_{135} \\
\textrm{N/A}
\end{bmatrix} \\
q &= \frac{Q}{I}\\
u &= \frac{U}{I}\\
\end{align}

In this Stokes formalism, $q$ and $u$ describe the fractional linear Stokes parameters and $I_{angle}$ denotes the intensity of light whose electric component propagates perpendicular to the specified angle from the reference frame of the camera.

\subsection{Detector Noise Propagation}
If the calibration of the camera is done is the best possible manner, the limiting factors of the polarimetric sensitivity will be the additive noise due to statistical fluctuations of the dark current and the Poissonian shot noise. 
Roussell et al.~have presented the propagation of these noise sources to the estimation of the stokes parameters in 2018. \cite{roussel_polarimetric_2018}

Here we quote the absolute lower limit of the variances on $I$, $q$, $u$ as the fundamental sensitivity that could be achieved by these kind of polarimeters: 

\begin{align}
    \sigma_{I} = \sigma_a+\sqrt{\frac{I}{2}} \\
    \sigma_{Q} = \sqrt{2}\sigma_a+\sqrt{I}  = \sqrt{2}\sigma_I \\
    \sigma_{U} = \sqrt{2}\sigma_a+\sqrt{I} = \sqrt{2}\sigma_I 
\end{align}

\begin{align}
    \sigma_{q} = \sqrt{(\frac{\partial q}{\partial I}\sigma_I)^2+\frac{\partial q}{\partial Q}\sigma_Q)^2} \\
    \sigma_{u} = \sqrt{(\frac{\partial u}{\partial I}\sigma_I)^2+\frac{\partial u}{\partial U}\sigma_U)^2}
\end{align}

\begin{align}
    \sigma_{q} = \frac{1}{I}\sqrt{q^2\sigma_I^2+\sigma_Q)^2} \\
    \sigma_{u} = \frac{1}{I}\sqrt{u^2\sigma_I^2+\sigma_U)^2}
\end{align}



Simplifying the expressions for $\sigma_q$ and $\sigma_u$ in terms of $\sigma_I$ then gives:

\begin{align}
    \sigma_{q} &= \frac{\sigma_I}{I}\sqrt{2+q^2} \\
    \sigma_{u} &= \frac{\sigma_I}{I}\sqrt{2+u^2}
\end{align}

For weakly polarized light,  approximate $I$ by the following equation:
\begin{equation}
    I \approx \frac{2S}{g}
\end{equation}

Where $S$ is the signal at a single detector pixel received in digital counts. Assuming only Poissonian shot noise, the expected noise on $q$ and $u$ will then be: 
\begin{equation}\label{eq:noise floor}
    \sigma_q \approx \frac{\sqrt{g}}{\sqrt{2S}} \text{  and  }
    \sigma_u \approx \frac{\sqrt{g}}{\sqrt{2S}}
\end{equation}

\begin{tcolorbox}[width=\linewidth, sharp corners=all, colback=white!95!black]
The fundamental photon noise, limiting the polarimetric sensitivity in $q$ and $u$ for a single superpixel in a single exposure, at user gain 0, unpolarized light, filling the pixel wells by 75\% is 8$\cdot10^{-3}$. If not limited by other effects, this random noise level can be suppressed by $\sqrt{N}$ for $N$ exposures to average over.
\end{tcolorbox}

\subsection{Dark Offset}
When no light is shining on the detector, there will still be a digital signal with both noisy and systematic characteristics. 
This common phenomenon is called the dark offset and mainly has two well-understood components. 
First is the readout offset, which we have assumed to be constant and independent of temperature, gain, or exposure time. 
Second and foremost is the signal due to the thermal generation of charge carriers. 
The electrons in the depletion layer of the photodiodes will produce a current, the strength of which is highly dependent on the temperature of the photodiodes. \cite{wang_noise_2008, widenhorn_temperature_2002} 
The total amount of dark electrons will go linearly with exposure time because of it being a constant current. 
The gain of the detector determines the multiplication factor between the number of electrons and the number of digital units registered. 
This multiplication factor cannot distinguish between photon-excited charge carriers and charge carriers excited due to a different mechanism. 
So the digital unwanted signal will be boosted in the same manner. 
Finally, the number of created charge carriers is exponentially related to the inverse of the temperature. 
Mathematically we can model it as follows:  

\begin{align}
    S_{\text{dark offset}} &= S_{\text{read offset}} + S_{\text{dark current}} \\
    S_{\text{dark current}} &= G(I_d*t) \\
    I_d &= e^{\zeta + \frac{\xi}{T}}  
\end{align}

where $S$ is the measured signal in digital units, $I_d$ is the dark current, $G()$ is the function that transfers electron into digital units as a function of the gain setting, $t$ is the exposure time in seconds, $T$ is the temperature in Kelvin. 
$\zeta$ and $\xi$ are two fitting parameters depending on the internal physical properties of the photodiodes.
$\xi$ is directly related to the energy gap of the semiconductor that the photodiode is made of and the dominant excitement mechanism. 
When $\xi$ is of the order of $\frac{E_g}{2k_B}$, the main excitement mechanism is thermal generation. 
However, when it is of the order of $\frac{E_g}{k_B}$, diffusion of the charge carriers in the neutral areas of the photodiode dominates. \cite{wang_noise_2008}

\begin{figure}
    \begin{subfigure}[b]{0.49\textwidth}
        \centering
        \includegraphics[width=\textwidth]{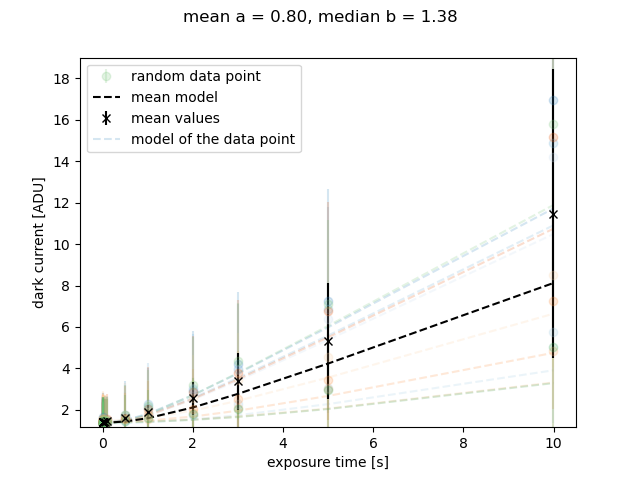}
        \caption{Dark offset with exposure time}
        \label{fig:dark vs. t}
    \end{subfigure}
    \begin{subfigure}[b]{0.49\textwidth}
        \centering
        \includegraphics[width=\textwidth]{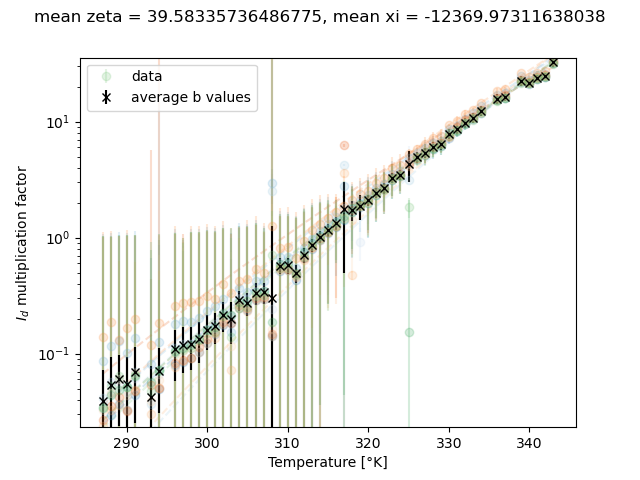}
        \caption{Dark offset with temperature}
        \label{fig:dark vs. T}
    \end{subfigure}
    \caption{Dependence of dark offset on exposure time and temperature.}
\end{figure}

\begin{figure}
    \centering
    \includegraphics[width=0.5\textwidth]{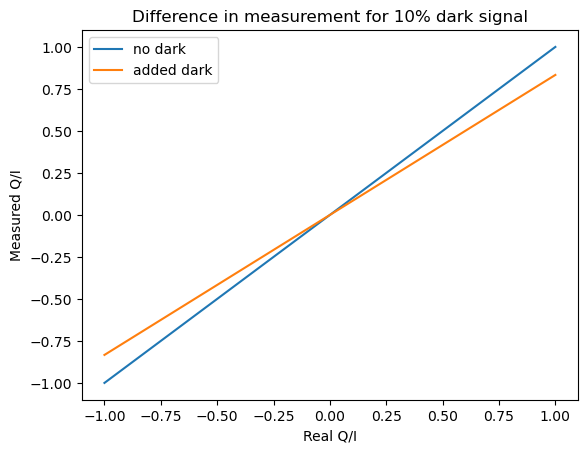}
    \caption{The simulated, measured $Q/I$ by a pixelated detector as a function of the real $Q/I$, when skewed due spurious dark signal of 10\% compared to ideal dark subtraction. }
    \label{fig:simulated_uniformdark}
\end{figure}

The polarimetric impact of neglecting or underestimating a spatially uniform dark offset is an induced skewing of the incoming polarimetric signal. 
In figure \ref{fig:simulated_uniformdark} we have simulated the impact of a spurious uniform dark offset of 10\%. It shows that the impact for weakly polarized light is low, but it becomes more relevant for strongly polarized sources. 
Mathematically, it can be easily shown that this skew is linear:

\begin{equation}
    \frac{q_{\text{measured}}}{q_{\text{real}}} = \frac{S_{\text{light}}}{S_{\text{light}}+2S_{\text{dark offset}}}
\end{equation}

This effect will be negligible for purposes with either very bright sources or mainly unpolarized sources. 
For faint, highly polarized sources it becomes very relevant. 

The main cause for a spurious dark signal is by not accounting for any temperature differences between the estimating the dark and the actual measurements. Due to the exponential relation between the temperature and the dark offset, the spurious dark signal can be substantial. From starting the camera to continuous operation, we have seen temperature differences of 20 degrees. Especially outside with changing light and wind conditions, fluctuations of the order of 10 degrees could happen over the course of 30 minutes.

\begin{tcolorbox}[width=\linewidth, sharp corners=all, colback=white!95!black]
Because of the exponential temperature dependency of the dark current, an underestimation of the dark offset is quickly done. To give a common example, when measuring at room temperature, taking 1-second exposures at a gain of 12, a two-degree difference in sensor temperature will result in a 5 digital unit underestimation of the dark offset. For 25\% full well light signal, this underestimation means the induced skew will be on the order of 1 percent for fully polarized light.
\end{tcolorbox} 

\begin{figure}
    \begin{subfigure}[b]{0.49\textwidth}
        \centering
        \includegraphics[width=\textwidth]{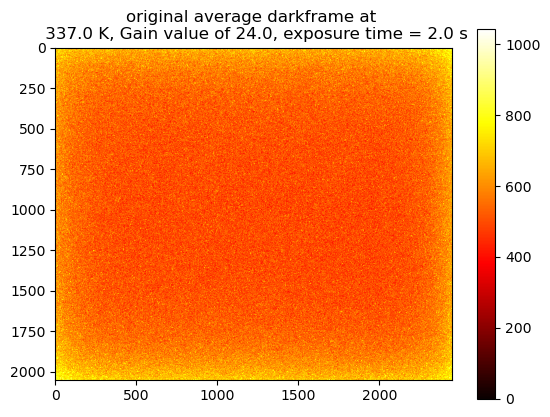}
        \caption{Real dark frame}
        \label{fig:darkreal}
    \end{subfigure}
 \begin{subfigure}[b]{0.49\textwidth}
    \centering
    \includegraphics[width=\textwidth]{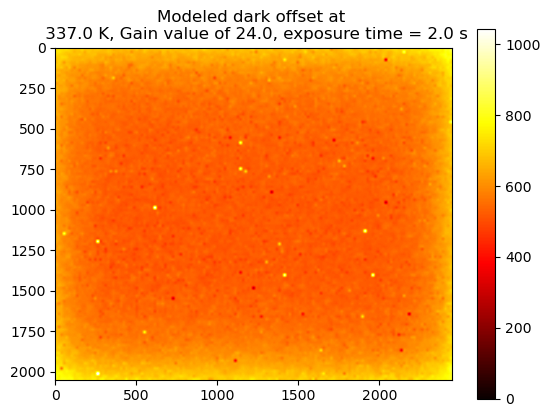}
    \caption{Modeled dark frame at a 16 coarser resolution}
    \label{fig:darkmodel}
\end{subfigure}
    
    \centering
    \caption{A dark frame for the LUCID camera at a device temperature of 337 Kelvin. Gain value was set to 24 and integration time set to 2 seconds}
\end{figure}

\begin{figure}
        \centering
        \includegraphics[width=\textwidth]{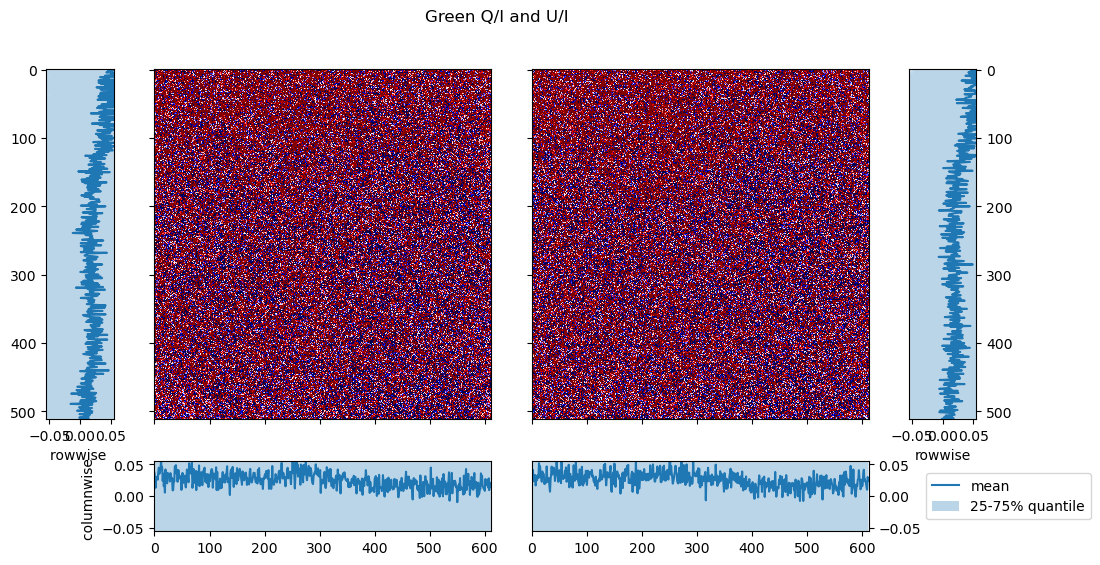}
        \caption{$Q/I$ and $U/I$ measured at all green pixels for a dark frame taken by the LUCID camera at a device temperature of 337 Kelvin. Gain value was set to 24 and integration time set to 2 seconds}
        \label{fig:darkreal_pol}
\end{figure}

\begin{figure}
    \centering
    \includegraphics[width=\textwidth]{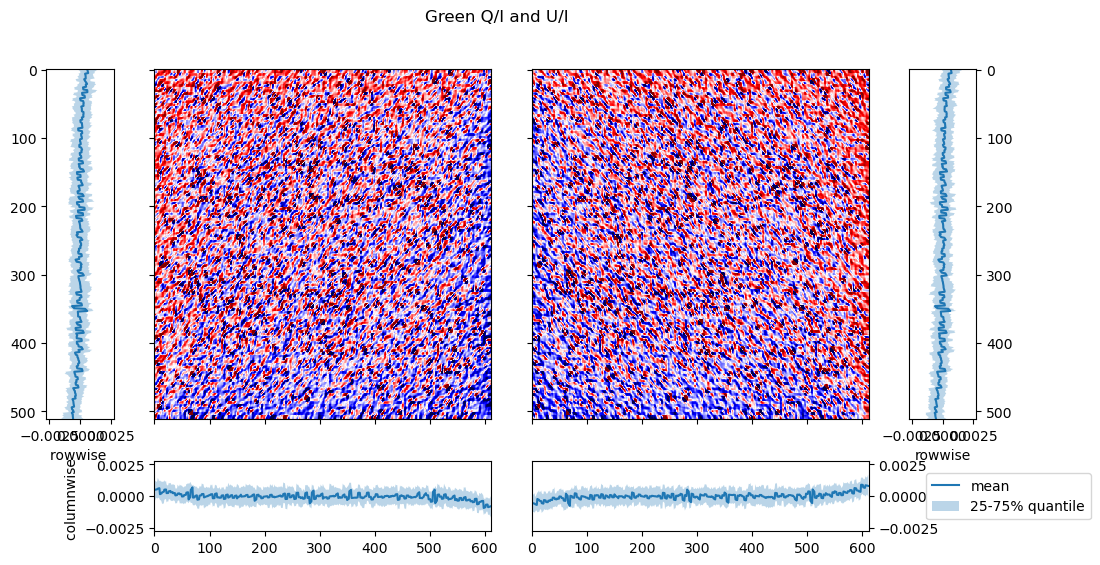}
    \caption{$Q/I$ an $U/I$ measured at all green pixels for the modeled dark frame of the LUCID camera at a 16 times coarser resolution, at a device temperature of 337 Kelvin, Gain value of 24 and an integration time of 2 seconds}
    \label{fig:darkmodel_pol}
\end{figure}

The dark offset is spatially varying over the detector. 
In all detectors, we see a glowing at the edges of the detector. 
This dark offset intensity gradient over the detector will be picked up as a polarimetric signal. 
In figures \ref{fig:darkreal}-\ref{fig:darkmodel_pol} we show some of the  worst cases dark frames, where this gradient is the strongest. 
Notice that the large-scale intensity gradient, as visualized by the modeled dark is of the order of $10^{-4}$ polarimetric offset.
However, the polarimetric offset using a single dark frame exposure is of the order of 1 percent, which is 2 orders of magnitude off. This shows the need for correct dark frame modeling for measurements of faint objects. 

The noise related to the read offset will be an additive noise per pixel. 
We assume that the noise of the dark offset is due to the poissonian nature of the dark current. 
One other argument for the use of a modeled dark instead of a single dark frame is the addition of the noise due to the dark current: 
\begin{equation}
    \sigma_{\text{dark corrected}} = \sqrt{2}\sigma_{\text{dark offset}}
\end{equation}

The effect on the polarimetric accuracy will be inversely proportional to the amount of light shining on the detector. 
Although the magnitude of the dark offset is linear with the exposure time, the strength of the gradient is not because it is a relative effect. 
To reduce the effect of this gradient for faint sources, longer exposures will benefit the measurement.


\begin{tcolorbox}[width=\linewidth, sharp corners=all, colback=white!95!black]
The offset due to the dark signal gradient will only form a problem when dark offset subtraction is done with a single value instead of a modeled dark frame in low-light conditions. However, even in these conditions it will be of the order of $10^{-4}$ at most. 
\end{tcolorbox}

\begin{figure}
    \centering
    \includegraphics[width=0.49\textwidth]{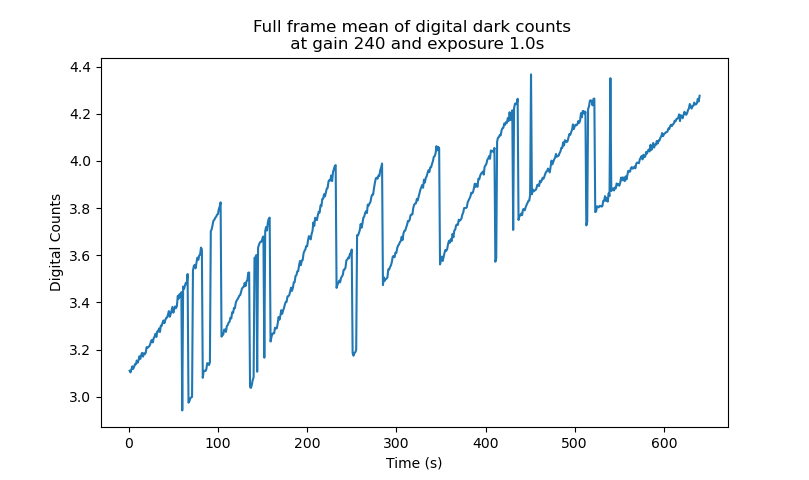}
    \caption{The mean digital count signal while the Thorlabs Kiralux camera slowly heated up from fridge- to room temperature and was being armed and disarmed between each subsequent acquisition. Since there is no way to read the sensor temperature of the detector through the Python SDK, time is used as an analogue for temperature and should be interpreted as 'time since the first image was captured.}
    \label{fig:thorlabs_darks}
\end{figure}

The discussion of the effects above holds under scrutiny for three of the four cameras. 
The exception is the Kiralux camera from Thorlabs. 
This camera performs an automatic dark subtraction that cannot be disabled either through the Python SDK or in Thorlabs' proprietary software. 
After initial testing, we suspect the automatic dark offset subtraction is a fixed number of digital counts for the whole sensor, such that the mean dark counts remain at approximately the same level. 
We could not rule out what exactly triggers the amount of correction is that applied. 
We suspect that the (amount of) dark correction is influenced by sensor temperature and the gain set by the user, but as can be seen in figure \ref{fig:thorlabs_darks}, the dark level can be made to chaotically oscillate by 1 digital count. 
In other words, as far as we can judge, there is no way to reliably and deterministically predict or model the amount of dark counts in the Thorlabs camera, making it unsuitable for high-accuracy polarimetric scientific endeavors.

\subsection{Non-linearity}\label{sec:linearity}
In all calibration schemes above, it is assumed that the CMOS sensors would exhibit exact linear behavior, i.e.~the signal would be linearly dependent on the exposure time and/or the photon flux. 
In this section we investigate the non-linearity to test this assumption by modeling its behaviour. 
The goal of this model is to calibrate the detector in such a way that we can reconstruct what the signal should have been were the detector perfectly linear.

To this end, we performed a series of measurements with each camera where the camera was placed in a collimated, unpolarized beam described in section \ref{sec:setup} and took exposures from the lowest possible exposure time up to the point where the entire sensor became saturated, at a given gain setting. 
With this dataset we can quantify how much the cameras deviate from linear behavior by evaluating how well we can fit a (non-)linear model to it. 

We assume that the signal we receive from our detector can be deconstructed into the following components:

\begin{equation}\label{eq:intensity-quadratic}
    S_{\text{raw}} = S_{\text{read offset}} + S_{c} + f(I^2(t)) \,.
\end{equation}

Here $S_{\text{raw}}$ represents the \emph{raw} signal of the detector, i.e. the signal/counts/intensity directly reported by the detector (this will also be the convention in the rest of this manuscript). 
$S_{\text{read offset}}$ represents the camera signal at zero photon flux and zero exposure time and $S_c$ represents the combined signal caused by the dark current and the photon flux. 
Finally,  $f(S^2(t))$ represents an unknown function that is quadratic in intensity, which we attribute to a non-linear effect in the amplifier circuitry of the detector. 
We assume this effect to be dominated by the in-situ signal acquired by the CMOS chip, independent of the physical brightness of the detected object.

We can rewrite the intensity equation to a form that depends explicitly on time:

\begin{equation}\label{eq:intensity-quadratic alternate}
S_{\text{raw}}(t) = S_{\text{read offset}} + G(I_\text{dark} + I_\text{light} )t + \frac{f(S^2(t))}{t^2}t^2 \,.
\end{equation}

Here $I_\text{dark}$ and $I_\text{light}$ represent the dark and light currents, respectively. 
As seen in section \ref{sec:setup}, the experimental setup for our calibration contains a light source that is sufficiently bright for us to make the assumption that $I_\text{dark}+I_\text{light}\approx I_\text{light}$.
Seeing that equation \ref{eq:intensity-quadratic alternate} has the form of a second-order polynomial, we can fit the model function $S=a+b\cdot t+c\cdot t^2$ to the data we acquired for each pixel independently. 
We can then relate the fit parameters directly to the desired physical quantities.

\begin{align}\label{eq:quadratic fit parames}
\begin{split}
    a &=  S_{\text{read offset}}\\
    b &= G(I_\text{light}) \\
    c &= \frac{f(S^2(t))}{t^2} 
\end{split}
\end{align}

It is important to note here explicitly that a choice in fitting strategy will influence the interpretation of the results later on. 
In performing the fitting, we have opted to optimize for the low signal intensity regime instead of the regime near pixel saturation. 
This will have the effect of our fit predicting sensor values larger than the saturation level for very high signal intensities.

\begin{figure}
    \centering
    \includegraphics[width=0.3\linewidth]{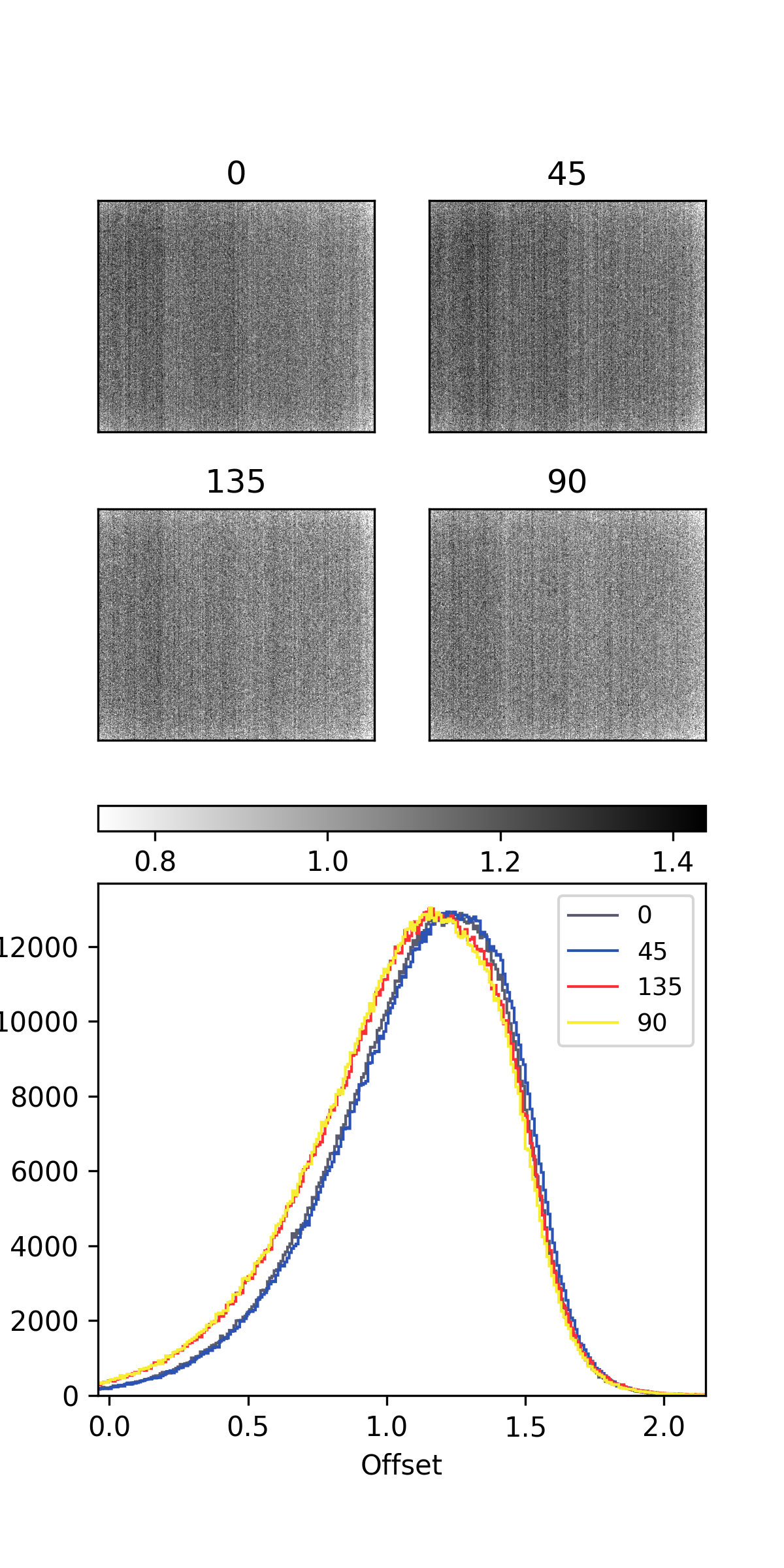}
    \includegraphics[width=0.3\linewidth]{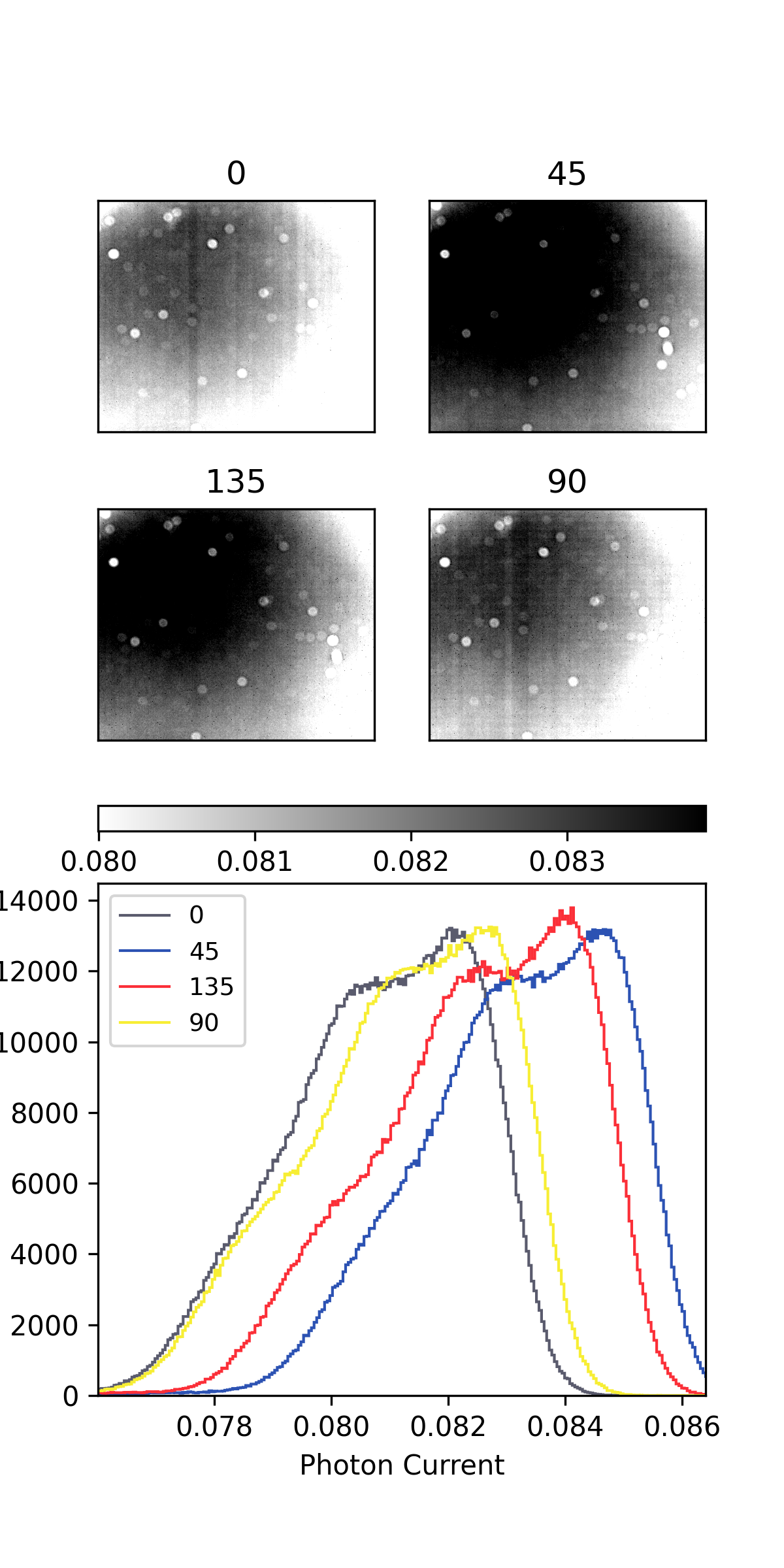}
    \includegraphics[width=0.3\linewidth]{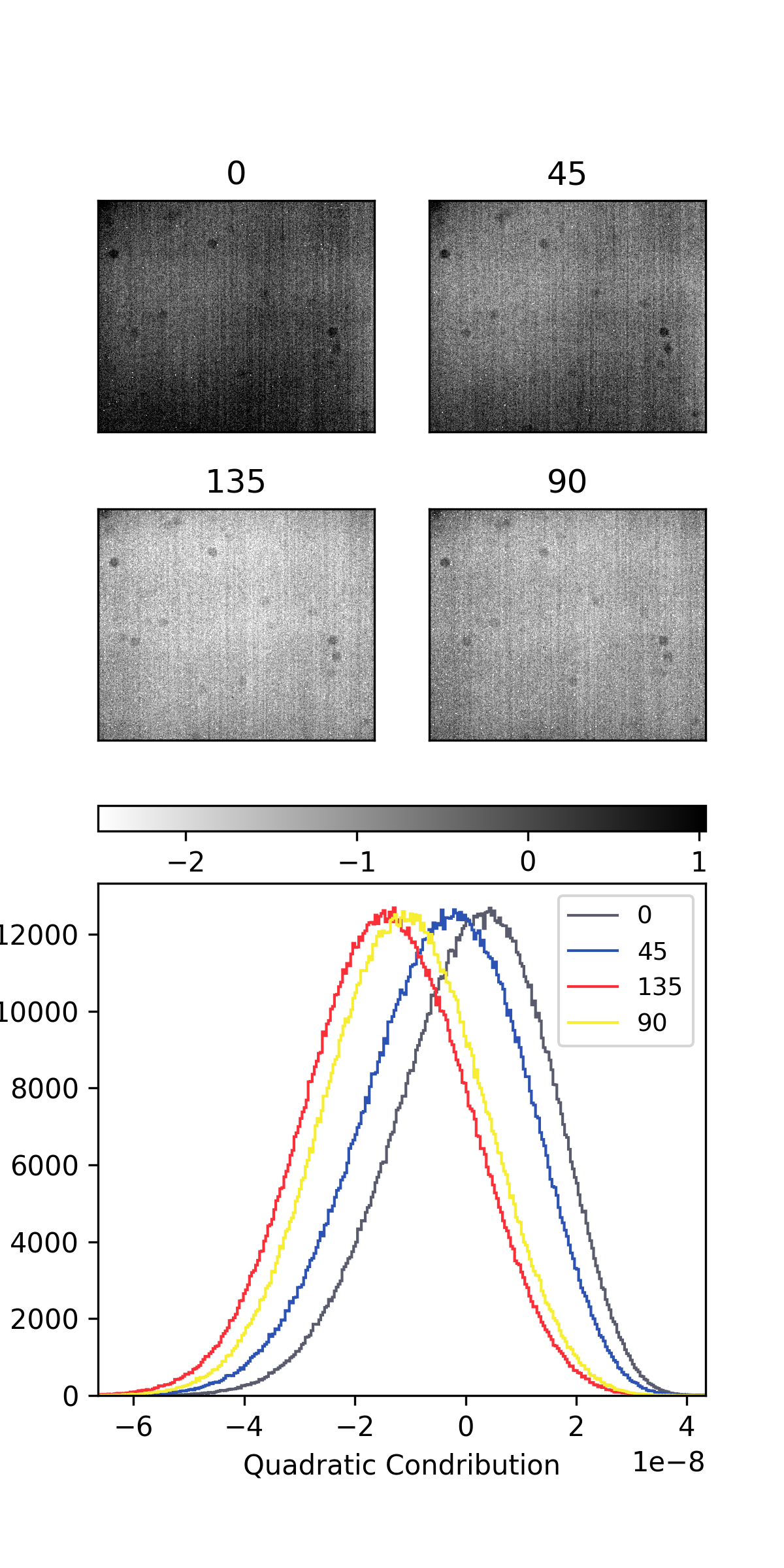}
    \caption{Fit parameters acquired per pixel at zero gain for the FLIR monochrome camera. The columns represent the three different fit parameters from equation \ref{eq:quadratic fit parames}. The top half shows a visual overview of the value of the fit parameters per pixel, where the sensor is split into its four polarization channels. The bottom half shows the same data in histogram form. }
    \label{fig:quadratic_fit_parameters}
\end{figure}

\begin{figure}
    \centering
    \includegraphics[width=0.7\linewidth]{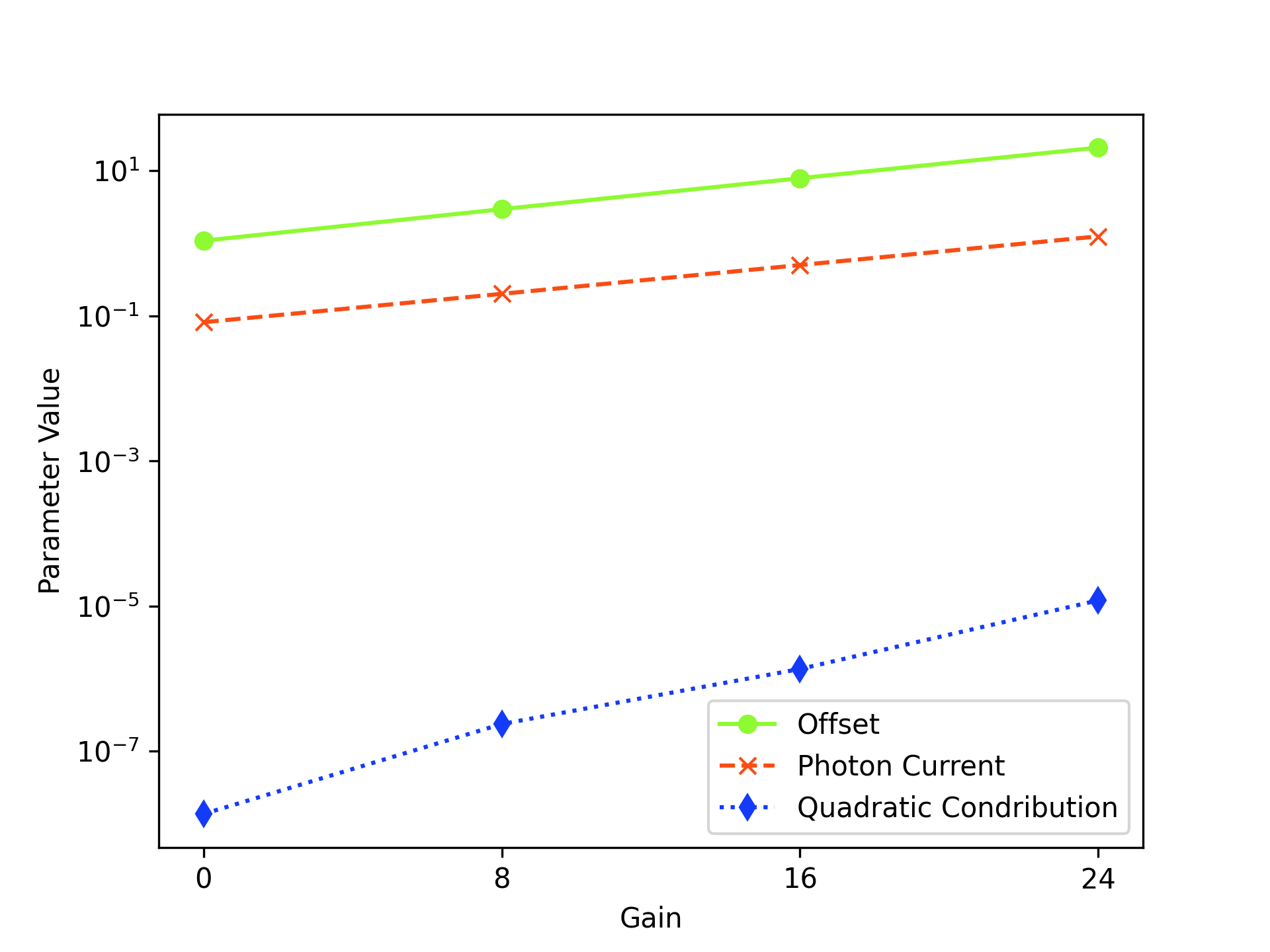}
    \caption{Values of the three fit parameters from equation \ref{eq:quadratic fit parames} as a function of sensor gain. Values presented here were acquired from the FLIR monochrome camera.}
    \label{fig:fitparames per gain}
\end{figure}

Figure \ref{fig:quadratic_fit_parameters} shows the results of one sensor-wide fit and how the parameters depend on the polarization channel of the sensors and figure \ref{fig:fitparames per gain} shows the mean values of the three parameters and how they scale with gain.  
We can surmise a few things from these two figures. 
Firstly, the linear component is typically many orders of magnitude larger than the quadratic component, indicating that the non-linear relevance is mostly limited to (very) low signal levels. 
The value of the offset is on the order of individual counts and corresponds to the lowest value the sensor typically outputs. 
Finally, we see that the photon current and the quadratic component depends to a significant degree on the orientation of the micro-polarizers. 
We can relate back to figure \ref{fig:darkreal_pol}, where we also see a polarimetric offset that is not related to the overall gradient. 
This combination suggests that this effect is due to the gain difference between the channels and not transmission and quantum efficiency differences which would only affect photon-induced electrons.  

Considering the results from the per-pixel quadratic intensity fits, there are two corrections we can introduce to enhance the polarimetric performance of our detector. 
This is a direct extension of the single-pixel calibration described by Powell and Gruev (2013) with the non-linearity calibration. \cite{powell_calibration_2013}
Firstly, we can remove the non-linear behavior and sensor offset to reconstruct the photonic signal. 
Given the values of the three fit parameters, we define a per-pixel correction procedure that transform any raw signal into a corrected signal. 
Using equations \ref{eq:intensity-quadratic} and \ref{eq:quadratic fit parames} we get 

\begin{equation}\label{eq:linearity correction equation}
S_{\text{cor}} = S_{\text{raw}} - a - c\hat{t}^2 \,,    
\end{equation}

where $\hat{t}$ is the time associated with the same raw signal intensity during the calibration measurement and is an easy proxy parameter to map all measurements back to the calibration measurement. 
This ‘associated $\hat{t}$’ can be calculated using the fit parameters of the calibration measurement:

\begin{align}
\begin{split}
\hat{t} &= \frac{-b + \sqrt{b^2 - 4c(a - S_{\text{raw}})}}{2c} \,.
\end{split}
\end{align}

\begin{figure}
    \centering
    \includegraphics[width=0.9\linewidth]{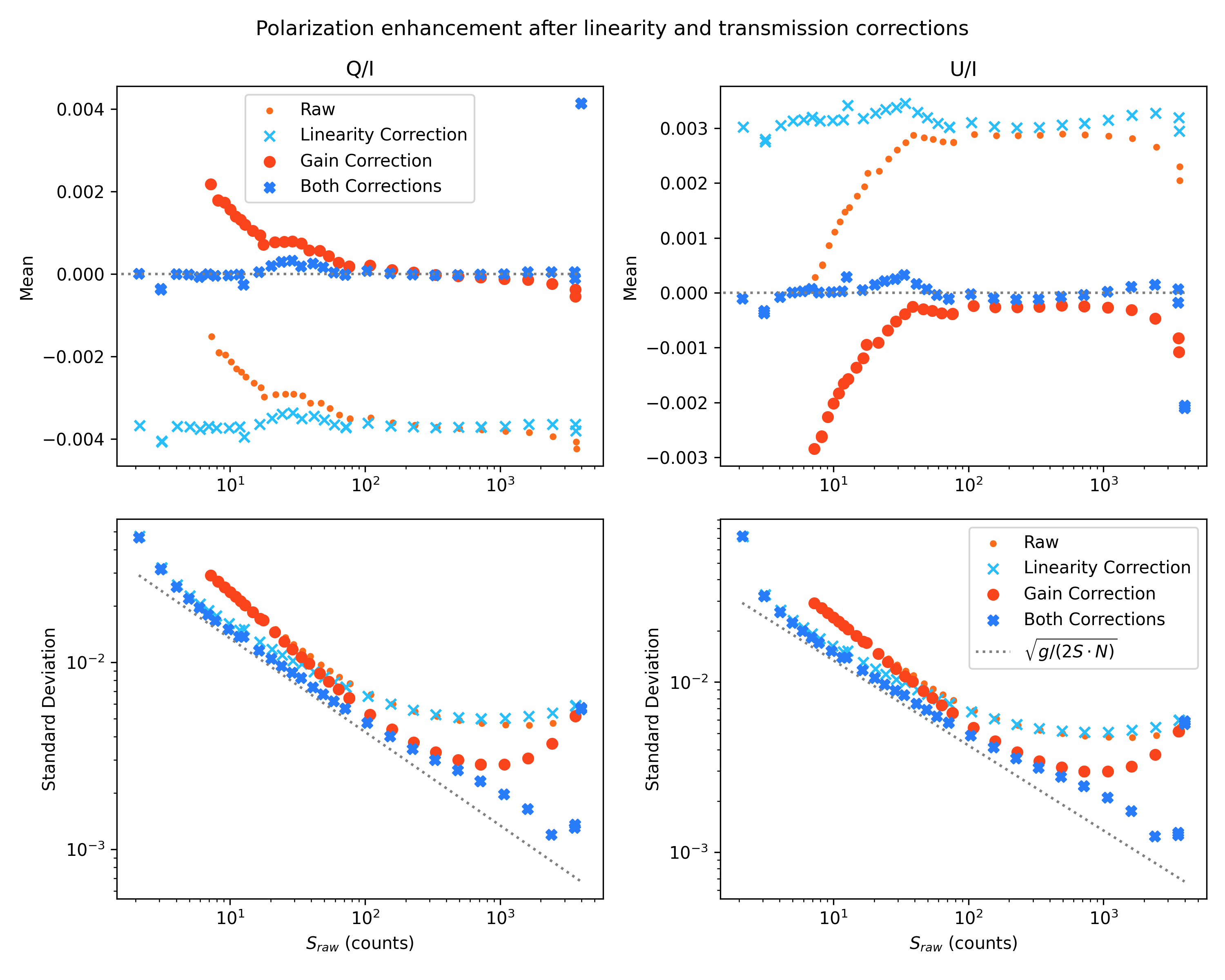}
    \caption{Enhancement of the polarization value as a function of mean signal intensity. The columns show $Q/I$ and $U/I$, with the top row representing the mean value over the whole sensor and the bottom representing the standard deviation from that mean. The dotted line in the top row represents the expected zero-signal, while the dotted line in the bottom plot represents the shot noise limit of the precision. Data acquired for the Thorlabs camera at zero gain with 100 frames per exposure time.}
    \label{fig:non-linearity enhancement per counts}
\end{figure}

If we reconsider the values of $q$ and $u$ for each exposure time during the calibration measurement, we can visualize the impact of the non-linearity correction as a function of signal intensity, as seen in figure \ref{fig:non-linearity enhancement per counts}. 
We clearly see that the linearity correction has a homogenizing effect in the low signal intensity regime, i.e. the polarimetric response - while not yet at the correct absolute value - is flattened over the entire signal range. 

\begin{tcolorbox}[width=\linewidth, sharp corners=all, colback=white!95!black]
    Correcting for non-linearity homogenizes the polarimetric signal at low signal intensities but does not increase the absolute accuracy of the polarimetric zero-point.

    If the gain correction (see next section) is applied \emph{without} a non-linearity correction, one can expect a mean residual error on the order of 0.1\% at signal levels below 100 counts.
\end{tcolorbox}

\subsection{Superpixel Differential Gains}\label{sec:gains}
As we noted in the previous section, there is a significant heterogeneity in gains between the four polarization channels (see figure \ref{fig:quadratic_fit_parameters}), which will clearly lead to an error in the measured values of polarization. 
Using the same fit parameters that gave us the non-linearity correction in the previous section, we can devise a correction strategy for these differential gains.

One path to correcting this effect is to homogenise the values of $b$ in equation \ref{eq:quadratic fit parames} for the entire sensor. 
However, this would risk the correction including the compound effects of both heterogeneous transmission and non-uniform illumination of our sensor during calibration. 
To mitigate the mixing with non-uniform illumination, we consider the scaling of this transmission correction per superpixel, assuming any non-uniformity of illumination is negligible on this scale. 
Note that this is still a single-pixel calibration since the interaction within the superpixel is neglected and the angle misalignment and diattenuation problems are not addressed in this way.

There is no unique way to correct for heterogeneous gain within a local region (superpixel). 
We opted for selecting the pixel with the lowest value of $b$ and scaling all four pixels to that value. 
This creates a \emph{gain correction map} with typical values $\in[1,2)$. 
This map is unique per camera and in essence scales $Q$ and $U$ to their correct value. 

Figure \ref{fig:non-linearity enhancement per counts} shows the effect of the gain correction, as well as the compound correction of both linearity and gain. 
The results show that the gain correction purely scales $q$ and $u$ towards zero and has no effect on the deviations at low signal intensities.
However, when we consider the compound correction, we get the best of both worlds: a flat polarization response centered closely around zero.

Furthermore, we can consider the spread of the values of $q$ and $u$ on the sensor (bottom half of figure \ref{fig:non-linearity enhancement per counts}). 
Primarily, independent on whether a correction is applied, we see that the standard deviation of the measured values scales by the factor $\sqrt{g/(2S\cdot N)}$, which is consistent with what we would expect as the noise limit (see equation \ref{eq:noise floor}). 
This implies that the spread can be further reduced by increasing the number of measurements $N$.
Secondarily, we see that the absolute spread of the data decreases by applying the gain correction and even further after the compound correction. 

\begin{figure}
    \centering
    \includegraphics[width=0.9\linewidth]{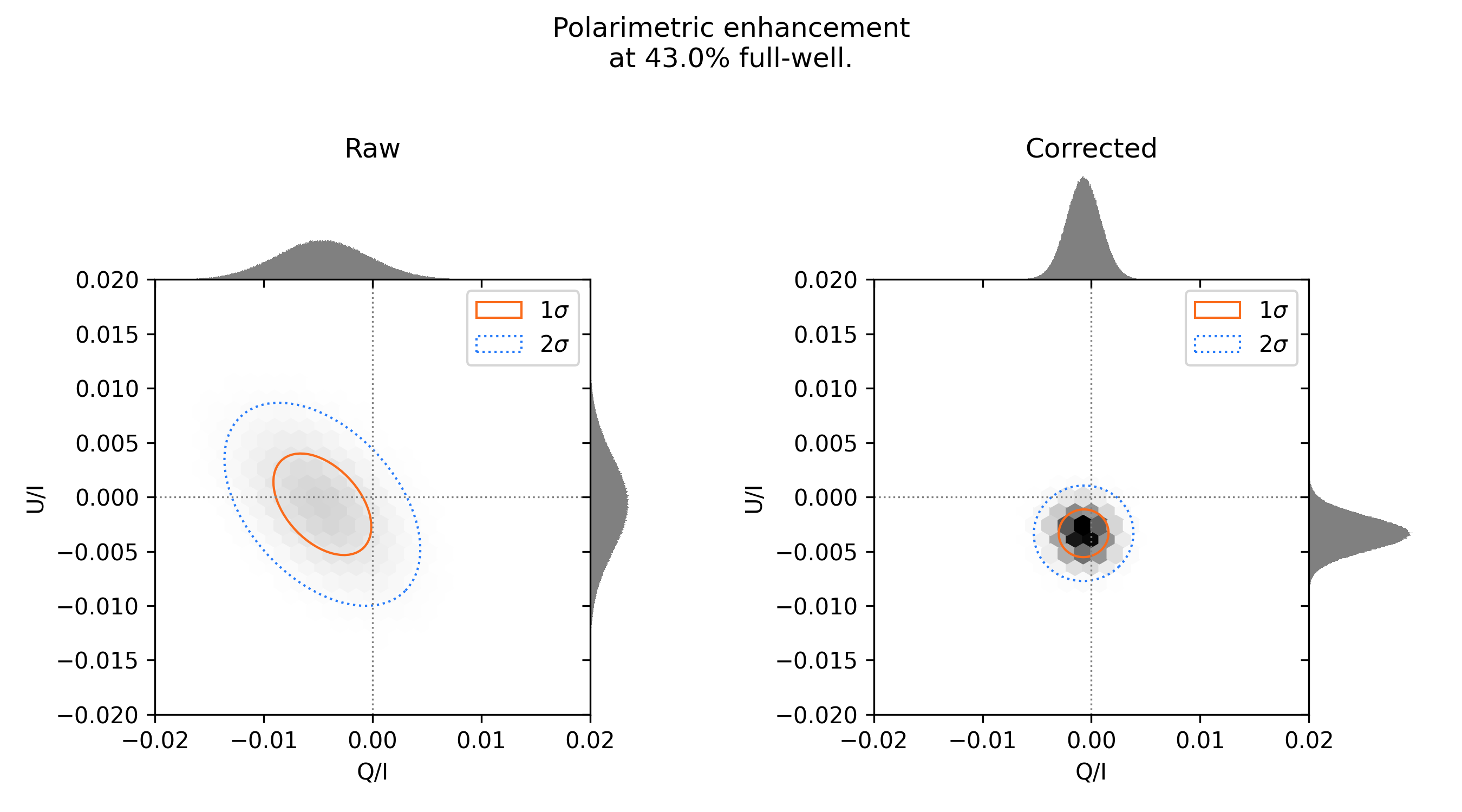}
    \caption{Polarimetric Enhancement of correcting for the non-linearity in the sensor and the pixel transmission differences per superpixel. Image obtained with the Thorlabs camera at zero gain and illumination such that the pixels are - on average - filled at 43\% full-well.}
    \label{fig:non-linearity enhancement}
\end{figure}

We validated the performance of the correction by applying it to a dataset of flat fields taken at a later point in time in the same experimental setup but with the camera rotated 90 degrees compared to the calibration measurements. 
The result can be seen in the 2D-histogram in figure \ref{fig:non-linearity enhancement}. 
As predicted, the histogram shows that the values of the superpixels become more centered around zero and vary less accross the sensor. 

\begin{tcolorbox}[width=\linewidth, sharp corners=all, colback=white!95!black]
    Gain non-uniformity will create an average error in the zero-point of $q$ and $u$ of the 0.4\%. By creating a sensor-specific gain correction map that re-scales the signal on a per-superpixel level, one can expect in the median that the values of $q$ and $u$ decrease by a factor of $3\cdot10^{-1}$ to 0.12\% and the values to be spread out 10\% less.
    
    The compound correction of gain and non-linearity can yield a median improvement in the values of $q$ and $u$ of $3\cdot10^{-2}$ to 0.012\% and an improvement in the spread of the values of 20\%.
\end{tcolorbox}

\subsection{Micropolarizer Misalignment}\label{sec:orientation of polarizers}
To evaluate the polarimetric performance of the cameras, we set out to track their intensities as a function of varying incoming linear polarization angle. 
We added a linear polarizer on a rotating stage to our collimated beam setup and measured the camera response for 36 equally spaced angles of rotation, as described in section \ref{sec:setup}.
We performed the measurement thrice with three different band-pass filters from the Thorlabs FBH series, centered around 450nm, 550nm and 650nm, all with their FWHM at 40nm. 

The signal of the cameras follows the Malus Law of transmission between two linear polarizers at mutual angle $\theta$:

\begin{equation}\label{eq:Malus law}
    I (\theta) = I_0 \cos^2(\theta - \phi) + c \,
\end{equation}

where $I_0$ is the signal amplitude, $\phi$ is an angle offset between the the rotating polarizer and the four different orientations of micro-polarizers on the detector, and $c$ is an offset of the zero-level. 
This expression can be rewritten into a form that is linear with a cosine- and a sine-component:

\begin{align}\label{eq:Malus law linear}
\begin{split}
    I (\theta) &= a + b \cos(2\theta) + c \sin(2\theta) \\
    a = \tfrac{1}{2} I_0 + c  \,,\quad 
    b &= \tfrac{1}{2} I_0 \cos(2\phi) \,,\quad
    c = \tfrac{1}{2} I_0 \sin(2\phi)  \,.
\end{split}
\end{align}

In this linear form, we can recognize the Mueller matrix elements $m_1=\tfrac{b}{a}$ and $m_2=\tfrac{c}{a}$, with which we can define the offset angle $\phi$:
\begin{equation}
    \phi = -\dfrac{1}{2}\arctan{\dfrac{m_2}{m_1}}\,.
    \label{eq:orientation_angle}
\end{equation}

Similarly to the procedure used in section \ref{sec:linearity}, we fit equation \ref{eq:Malus law linear} to each pixel individually and use that to create a map of the extinction ratio and angle offsets. 
We will discuss the angle offsets here and the extinction ratio in the next section.

The fitting was done on data without first correcting for non-linearity and transmission differences (see sections \ref{sec:linearity} and \ref{sec:gains}). 
Due to the design of the detectors - including microlenses to focus the incoming light on the polarizer - we assume the cross-talk between pixels to be negligible.
For the values of $\phi$ there is a freedom of reference frame. We choose to define the zero degree axis at the median value of theta of the pixels corresponding to zero degrees as defined by the manufacturer.  

\begin{figure}
    \centering
    \includegraphics[width=0.9\linewidth]{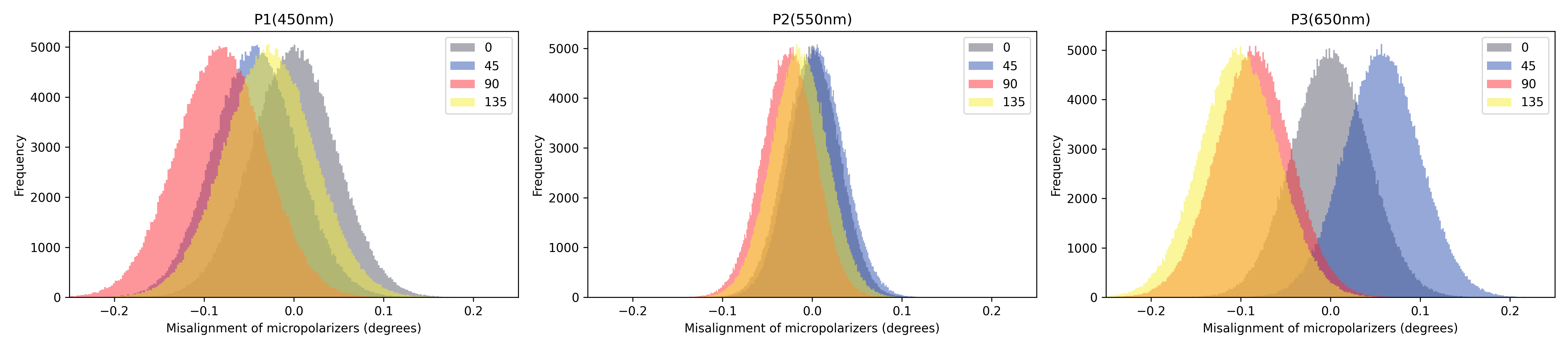}
    \caption{Histogram of micropolarizer misalignments. The zero point is defined as the position of the median of the pixels corresponding to the zero degree micropolarizers. The three columns correspond to results acquired with light filtered by three different band-pass filters. Data acquired for the LUCID RGB camera.}
    \label{fig:polarizer misalignment}
\end{figure}

Figure \ref{fig:polarizer misalignment} shows the results of the fitting process and the calculated values of $\phi$ per pixel. 
We can conclude two facts from this figure. 
Firstly, the amount of mutual misalignment appears to be on the order of 0.1 degrees. 
Secondly - and oddly enough - the amount and direction of misalignment appears to depend on the wavelength of light with which we investigate. 
We hypothesize that the root of this chromatic effect lies in the (orientation of) the color filters in our experimental setup. 
It is possible that these filters introduce a very slight retardance along one of its axes, shifting the polarization angle of the incoming light. 

As we cannot state with absolute certainty how much of the misalignment effects is caused by the camera's and how much by our experimental setup, it is important to note that any calibration that follows from this measurement needs to be treated with increased scrutiny, lest we calibrate our specific setup instead of the camera.
For now, we will continue on and consider mathematically what the impact a misalignment effect \emph{could} be in terms of polarimetric performance.

We can describe each micropolarizer as a Mueller matrix with perfect diattenuation with the general description

\begin{equation}
M = \frac{1}{2} \begin{bmatrix}
1 & cos(2\phi) & sin(2\phi) & 0 \\
cos(2\phi) & cos^2(2\phi) & cos(2\phi)sin(2\phi) & 0 \\
sin(2\phi) & cos(2\phi)sin(2\phi) & sin^2(2\phi) & 0 \\
0 & 0 & 0 & 0 
\end{bmatrix} 
\end{equation}

The intensity measured on each pixel will then be:

\begin{equation}
    I_{\phi} = \frac{1}{2}(I_{tot}+cos(2\phi)Q+sin(2\phi)U)\label{eq:polarizer_intensity}
\end{equation}

Any misalignment will produce a new Mueller matrix and will need a different solution to demodulate for $I$, $Q$, and $U$. 

\begin{figure}
    \centering
    \includegraphics[width=0.95\textwidth]{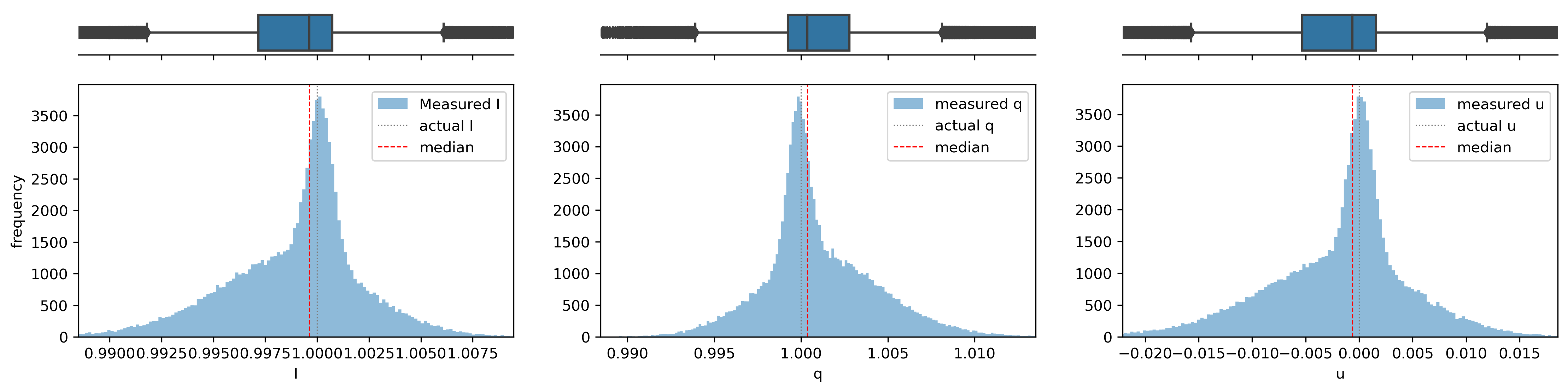}
    \caption{Retrieved I, q and u for 100000 random superpixels from the LUCID camera when illuminated with fully polarized light at 0 degrees.}
    \label{fig:misalignment_q}
\end{figure}

\begin{figure}
    \centering
    \includegraphics[width=0.95\textwidth]{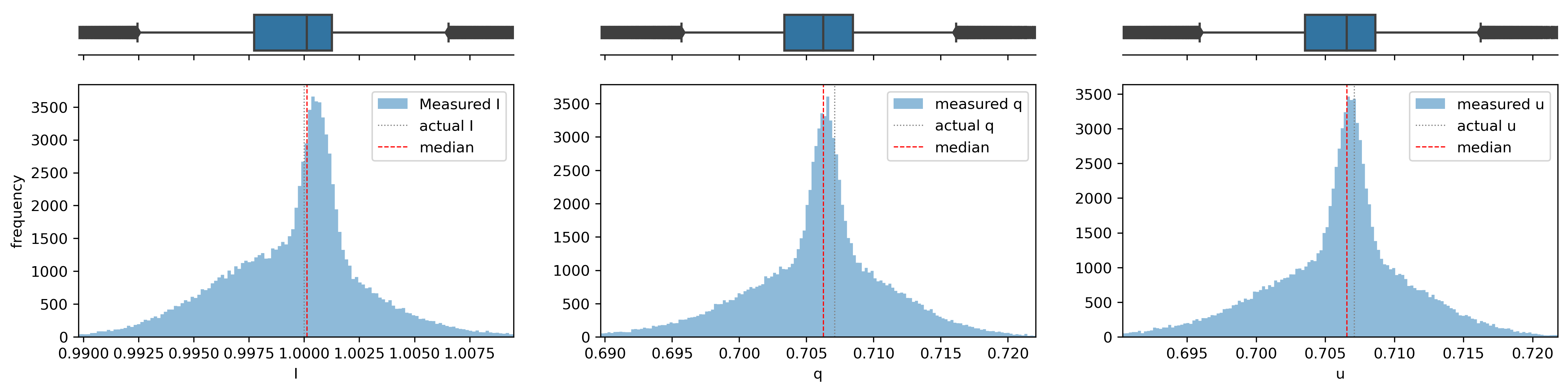}
    \caption{Retrieved I, q and u for 100000 random superpixels from the LUCID camera when illuminated with fully polarized light at 22.5 degrees.}
    \label{fig:misalignment_qu}
\end{figure}

To illustrate the impact of the pure misalignment on the polarimetric accuracy, we have mathematically propagated fully polarized light with through 100000 random superpixels sampled from the first panel in figure \ref{fig:polarizer misalignment}. For 0 degrees, the result can be seen in figure \ref{fig:misalignment_q}, and in figure \ref{fig:misalignment_qu} the angle of polarization is 22.5 degrees. Although the median measured intensity approximates the induced polarization well, the spread is of the order of $5*10^-3$, with long tails. 

\begin{tcolorbox}[width=\linewidth, sharp corners=all, colback=white!95!black]
When the misalignment is not corrected, the mistake in q and u for a single superpixel is likely to be 0.5 percent off for fully polarized light.  
\end{tcolorbox}

To properly correct the misalignment effect, one should make use of the "superpixel calibration" technique as described by Powell and Gruev in 2013. \cite{powell_calibration_2013}
However, this technique is focused on Stokes demodulation in the image plane, as the correction depends on the local intensity and polarization signals.
However, this demodulation is fundamentally limited by aliasing from the intensity structure across each superpixel.
In pure Fourier-plane demodulation methods, the aliasing can be mitigated, but only for many (super)pixels at the same time. 
For a fully reliable pixel-by-pixel calibration correction of the micropolarizer properties, an iterative algorithm needs to be developed.

\subsection{Extinction Ratio of Micropolarizers}
Continuing with the fit described in the previous section, we can calculate the extinction ratios of each pixel using equation using the ratio's of Mueller Matrix elements that we found from equation \ref{eq:Malus law linear}.

\begin{equation}
    \epsilon = \dfrac{1+\sqrt{m_1^2+m_2^2}}{1-\sqrt{m_1^2+m_2^2}}\,,
    \label{eq:ext_ratio}
\end{equation}

\begin{figure}
    \centering
    \includegraphics[width=1.1\linewidth]{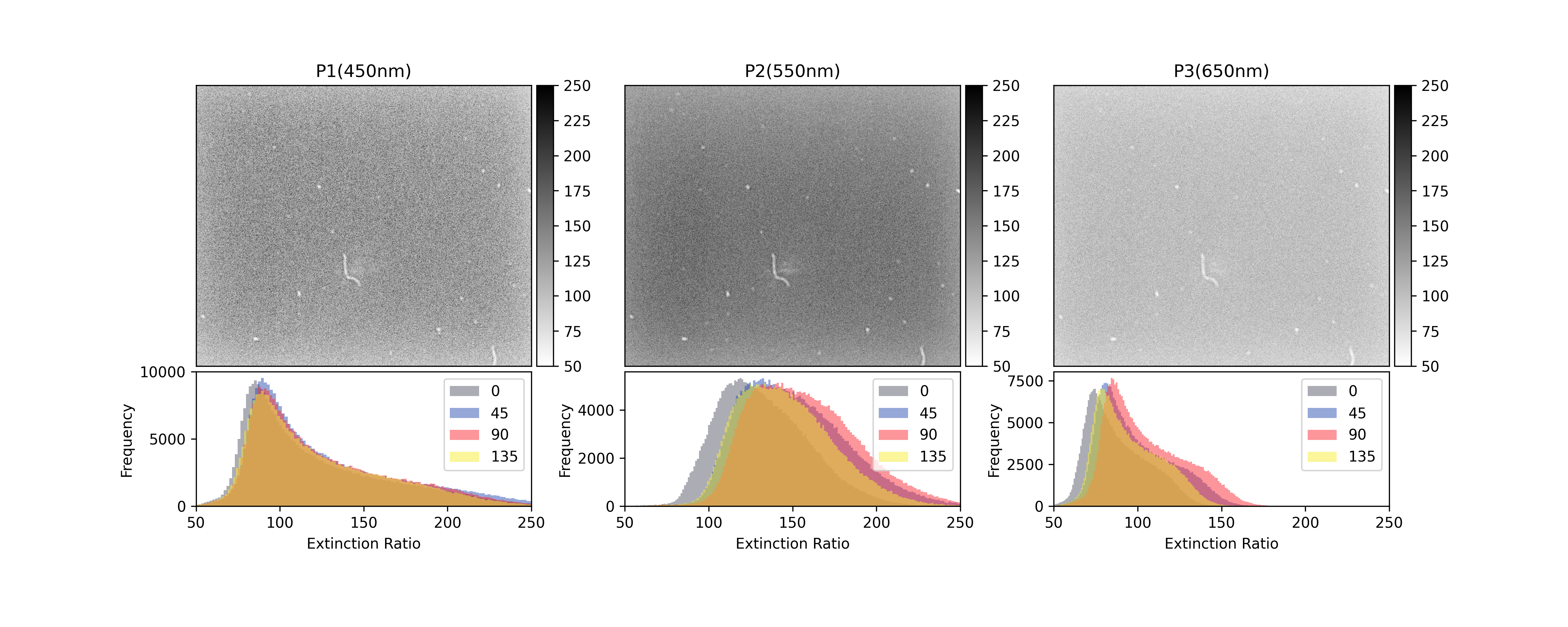}
    \caption{Extinction ratio of the detector on the LUCID RGB camera.}
    \label{fig:extinction ratio}
\end{figure}

Figure \ref{fig:extinction ratio} shows these values for one of the cameras, again split per broadband filter color. 

The measurements show that the value of the extinction ratios differs depending on the wavelength investigated. 
This chromatic dependency can be explained by the heterogeneous quantum efficiency of the CMOS sensors across the visible wavelength range.
However, we do not have a clear explanation of the difference in  the shapes of the distributions with the different wavelength bands. We expect a normal distribution of the diattenuation at all regimes, which would convert into a skewed distribution similar to the distributions seen in the P1 450 nm case (left most figure of figure \ref{fig:extinction ratio}). 
We suspect that there might be two gaussian profiles in the diattenuation distributions that become more distinct at higher wavelengths, where the red pixels are more dominant and general extinction ratios lower. 
We can confirm the conclusions of Hagen et al. 2019 where they note the importance of spatially characterizing the extinction ratio compared to only noting the mean value, which is a bad metric for quantifying the distribution compared to the median or the mode. \cite{hagen_calibration_2019}

A misestimation of the extinction ratios leads to an underestimation of the linear polarization signal. This skew is linear with the degree of linear polarization as was the case for the spurious dark offset. 

\begin{tcolorbox}[width=\linewidth, sharp corners=all, colback=white!95!black]
For fully polarized light, extinction ratios of 100 will result in 1 percent less $q$ and/or $u$ determined. 
\end{tcolorbox}

\section{System Level Analysis}
\subsection{Worst Offenders}
To summarize the conclusions of above we note its expected significance for every effect under four common situations in table \ref{tab: summary}.
\begin{table}[]
    \centering
    \caption{Summary of the effects at different measurement conditions. Every effect that is relevant for a certain level of illumination ($S_\text{raw}$) and Degree of Linear Polarization (DoLP) is marked with an X}
    \label{tab: summary}
\begin{tabular}{|c|c|c|c|c|}
\hline
\textbf{DoLP}  & $\sim 1$ & $\sim 0$ & $\sim 1$ & $\sim 0$ \\
\textbf{$S_{\text{raw}}$}                & $\mathcal{O}(10)$   & $\mathcal{O}(10)$   & $\mathcal{O}(1000)$ & $\mathcal{O}(1000)$ \\ \hline
Uniform dark offset             & X                     &                       &                       &                       \\
Spatially varying dark offset   &                       &                       &                       &                       \\
transmission differences        &                       & X                     & X                     & X                     \\
Non-linearity                   & X                     & X                     &                       &                       \\
Misalignment of micropolarizers & X                     &                       & X                     &                       \\
non-ideal Extinction ratios     &                       &                       & X                     &                       \\ \hline
\end{tabular}
\end{table}

\subsection{Polarimetric Accuracy}
The main inherent limitation to the polarimetric accuracy is the pixel gain variations within each superpixel.
We find that we can correct the pixel gain non-uniformity to 0.0025 absolute accuracy, as can be seen in figure \ref{fig:non-linearity enhancement}. 
Darks can corrected down to intrinsic noise level in all cameras when taking note of the sensor temperature. An exception to this is the Thorlabs camera where the automatic dark subtraction will be the main source of polarimetric inaccuracy for highly polarized sources.
Micropolarizer misalignment and extinction ratios can be better corrected than the pixel gain non-uniformity and are not the main contribution to systematic offsets, particularly for weak polarization signals. We did not see major differences between the cameras for the pixel-differential gain, misalignment or extinction ratios, so do not expect a difference in possible accuracy. Finally, we did not see any indication of temporal degradation of any effect, but this would require another study with set revisit times.   

\subsection{Polarimetric Sensitivity}
We find that all uncertainty effects can be calibrated close to the fundamental noise level until bright illumination. However, with the uncertainties in the calibration fitting one introduces an additional random noise. 
The main random noise level is still predominantly due to shot noise when dealing with brighter sources and the poissonian noise of the dark offset for fainter sources. 
In this case, more measurements are essential to push the overall polarimetric uncertainty of a single pixel to sub-percent level. 
To give an indication: for measurements at 75\% full-well illumination and a user gain of 16, approximately 326 measurements are needed for 0.001 polarimetric sensitivity in determining $q$ and $u$ for a single pixel.

\section{Addition of a rotating halfwave plate}\label{sec:HWP}
\begin{figure}[h]
	\centering
	\includegraphics[trim={1.7cm 3.9cm 1.3cm 3.3cm}, clip,width=\textwidth]{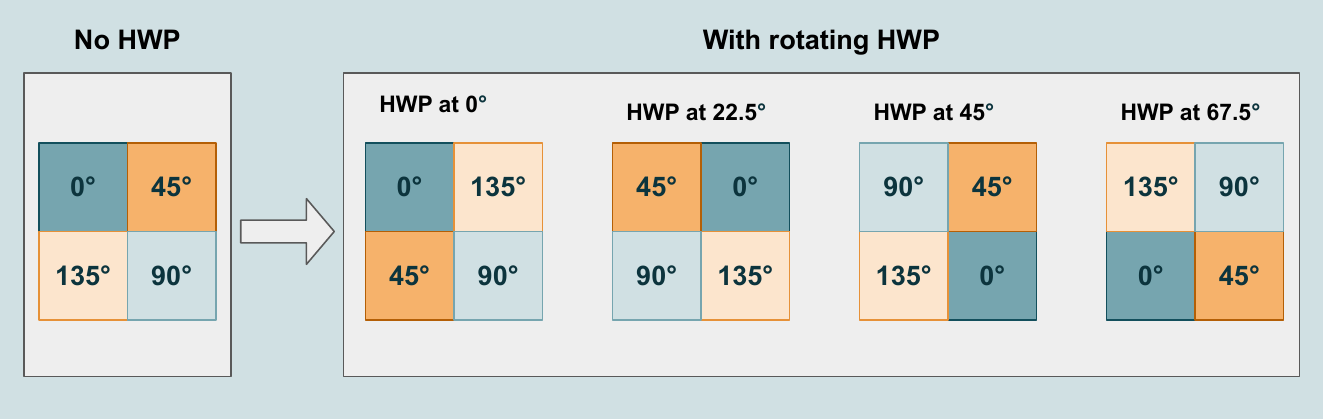}
	\caption{On the left-hand side a DoFP super-pixel is shown, with four distinct micro-polarizer angles. On the right-hand side, the same pixel is shown with an HWP at four different angles, changing the incoming polarization angle that will reach the detector pixel.}
	\label{fig:hwp_angles_demo}
\end{figure}

As an alternative to a full calibration of the DOFP cameras, we explored the potential of a demodulation method that relies on a combination of spatial and temporal modulation of the signal, through the addition of a rotating half-wave plate (HWP) to the set-up. 
The HWP temporally modulates the signal at four times its rotational angle, while effects that occur at the detector stay mostly constant.
Our expectation is that certain systematic errors that are introduced by the detector can be reduced or removed entirely by combining measurements at different HWP angles, at the expense of a reduced time resolution, but possibly gaining spatial resolution.\cite{vaughn_spatio-temporal_2019, li_optimal_2019}
This would be analogous to the so-called "dual-beam exchange" implementation that is prevalent in sensitive astronomical polarimetry.\cite{bagnulo_stellar_2009, snik_astronomical_2013} 
Variations on this approach have been implemented with a quarter-wave plate for full-Stokes polarimetric imaging\cite{shibata_robust_2019, shibata_video-rate_2019}, and with a vortex retarder\cite{li_vortex_2024}.\\
For this exploration of the topic, we chose a simple measuring sequence in which the desired measurements are done with the HWP at each of the angles $\beta\in(0^\circ, 22.5^\circ, 45^\circ, 67.5^\circ)$. 
In this way, the functions of which pixel analyzes which AoLP effectively get switched around, as can be seen in figure \ref{fig:hwp_angles_demo}. 
On the left, A regular super-pixel with four polarization filters is shown, with the angles representing the fixed angles of the micro-filters. 
To the right of the arrow, four situations are shown, where the HWP changes which polarization state is analyzed by each pixel, as depicted by the angles inside it.

\subsection{Theoretical Description}\label{sec: hwp theory}
When including a rotating HWP, the expression for measured intensity as given in equation \ref{eq:polarizer_intensity} has to be updated by combining the Mueller matrix for a rotated HWP and a rotated polarizer, both with their fast axis aligned with Stokes $Q$ at $0^\circ$. The intensity for general polarizer angle $\theta$ and general HWP angle $\beta$ is:
\begin{equation}
    I'_\text{det}=\frac{1}{2}(I_\text{in}+Q_\text{in}[\cos{2\theta}\cos{4\beta}+\sin{2\theta}\sin{4\beta}]+U_\text{in}[\cos{2\theta}\sin{4\beta}-\sin{2\theta}\cos{4\beta}]),\label{eq:I_det_for_general_hwp_and_pol_angles}
\end{equation}
By measuring at four HWP angles ($\beta\in(0^\circ, 22.5^\circ, 45^\circ, 67.5^\circ)$), and four fixed polarizer angles per super-pixel ($\theta\in(0^\circ, 45^\circ, 90^\circ, 135^\circ)$), we obtain sixteen variations on this expression, to which we can add our expected pixel-specific errors. The first error that we add is a transmission factor, $T_\theta$. This transmission factor is specific for each individual pixel. This factor represents every effect that scales with the total signal, independent of its polarization state. This includes differential gains of the pixels, conversion gain of the ADC, and (nonlinear) response of the photodiodes. Even though these effects operate on different scales, they are expected to be constant for a specific set of camera settings, and independent of the polarization state of incoming light. For this reason, they can be included in one multiplication factor. 
The second error we include is an offset that is completely independent of the photon signal which includes read offset and the dark current. Including this as a constant pixel-specific addition requires that the dark current stays constant, which means a constant sensor temperature. We will denote this term by $D_\theta$. 
The third error is the small angle offset of the microgrid polarizers, as discussed in  section \ref{sec:orientation of polarizers}. As the on-pixel intensity modulates with twice the polarizer angle, it also modulates with twice the angle offset. In equation \ref{eq:I_det_for_general_hwp_and_pol_angles} we can replace the $\sin{2\theta}$ and $\cos{2\theta}$ with an expression for the ideal angles $\theta$ with a small offset $\alpha$ in the small angle approximation:
\begin{align}
    \sin{(2\theta+2\alpha)}&\approx\sin{2\theta}+\cos{2\theta}\cdot2\alpha\\
    \cos{(2\theta+2\alpha)}&\approx\cos{2\theta}+\sin{2\theta}\cdot2\alpha
\end{align}
%
Lastly, we incorporate an expression for the error that arises as a result of the diattenuation imperfections of the polarizers. In this case, the transmission of polarization parallel to the fast axis can be denoted by $t_x$, and the transmission of polarization orthogonally to it by $t_y$. The $I\rightarrow I$ transmission gets a factor $t_x^2+t_y^2\equiv a$, and $Q\rightarrow I$ and $U\rightarrow I$ get a factor $t_x^2-t_y^2\equiv b$, which can both differ from pixel to pixel.

As an example, two HWP angles for the polarizer angle $\theta=0^\circ$ are implemented, $\beta=0^\circ, 45^\circ$:
\begin{align}
    I'_{0}(\beta=0^\circ)&=\frac{1}{2}T_0[a_0\cdot I_\text{in}+b_0\cdot(Q_\text{in}-2\alpha U_\text{in})]+D_0\\
    I'_{0}(\beta=45^\circ)&=\frac{1}{2}T_0[a_0\cdot I_\text{in}-b_0\cdot(Q_\text{in}-2\alpha U_\text{in})]+D_0\label{eq:App_some_errors_pol=0}
\end{align}
By linearly combining these two intensities, we can retrieve the stokes parameters. This holds in general for all orientations of the micropolarizers. For every single orientation within the super-pixel, $\theta$, a combination of two HWP angles can be used to reveal $Q$ and $U$ respectively. 
This can be represented by the generalized equations:
\begin{align}
    \hat{Q}_\theta&= I'_\theta(\beta_{\theta,0})-I'_\theta(\beta_{\theta,2})=b_\theta T_\theta\cdot[Q-2\alpha_\theta U]\\ 
    \hat{U}_\theta&= I'_\theta(\beta_{\theta,1})-I'_\theta(\beta_{\theta,3})=b_\theta T_\theta\cdot[U+2\alpha_\theta Q]\\
    \hat{I}_\theta&= I'_\theta(\beta_{\theta,0})-I'_\theta(\beta_{\theta,2})\\
    &= I'_\theta(\beta_{\theta,1})-I'_\theta(\beta_{\theta,3})=a_\theta T_\theta\cdot I+2D_\theta\nonumber
\end{align}
where $\beta_{\theta,i}=\text{mod}(\frac{\theta}{2}+i\cdot22.5^\circ, 90^\circ)$.
Consequently, we can write down the fractional Stokes parameters, as found through temporal demodulation for one pixel:
\begin{align}
    \hat{q}_\theta&=\frac{a_\theta}{b_\theta}\cdot\frac{Q-2\alpha_\theta U}{I+2D_\theta/T_\theta}\\
    \hat{u}_\theta&=\frac{a_\theta}{b_\theta}\cdot\frac{U+2\alpha_\theta Q}{I+2D_\theta/T_\theta}
\end{align}
These equations show that it is still very important to perform a proper dark subtraction. 
However, the effect of differential gain factors drops out of the equation. 
Some loss of signal occurs due to the diattenuation of the microgrid polarizers, as well as an offset as a result of the polarizer angles. Both these effects will get averaged over for each superpixel to retrieve the best estimate of our fractional linear stokes parameters. 
The average $\hat{q}$ and $\hat{u}$ can be computed as:
\begin{align}
    \hat{q}&=\frac{1}{4}[\hat{q}_0+\hat{q}_{45}+\hat{q}_{90}+\hat{q}_{135}]\\
    &=\frac{\hat b}{\hat a}\frac{2 Q_\text{real}-[\alpha_{0}+\alpha_{45}+\alpha_{90}+\alpha_{135}]\cdot U_\text{real}}{2 I_\text{real}}
\end{align}
%
Although these equations are only a first order approximation, they provide some insight in the expected effects. 
However, in real measurements, fluctuations of parameters or in brightness of the observed target might affect the results, and depend on the application that the sensor is used for. 
We expect an improvement in both polarimetric accuracy and sensitivity, mainly due to the transmission/gain terms dropping out completely. Dark frames have to be subtracted always to get a good result. The effect of decreased polarimetric sensitivity due to a non-ideal diattenuation will still occur, but any potential false polarization signal created by subtracting intensities from two different pixels will be removed. Polarizer misalignment is averaged over all four pixels, which is expected to provide some improvement in polarimetric performance, as we observed a dependence of angle offset to the expected pixel angle (see figure \ref{fig:polarizer misalignment}).

\subsection{Polarimetric Accuracy with a Halfwave Plate}\label{sec: hwp accuracy}
\begin{figure}[h]
	\centering
	\includegraphics[width=\textwidth]{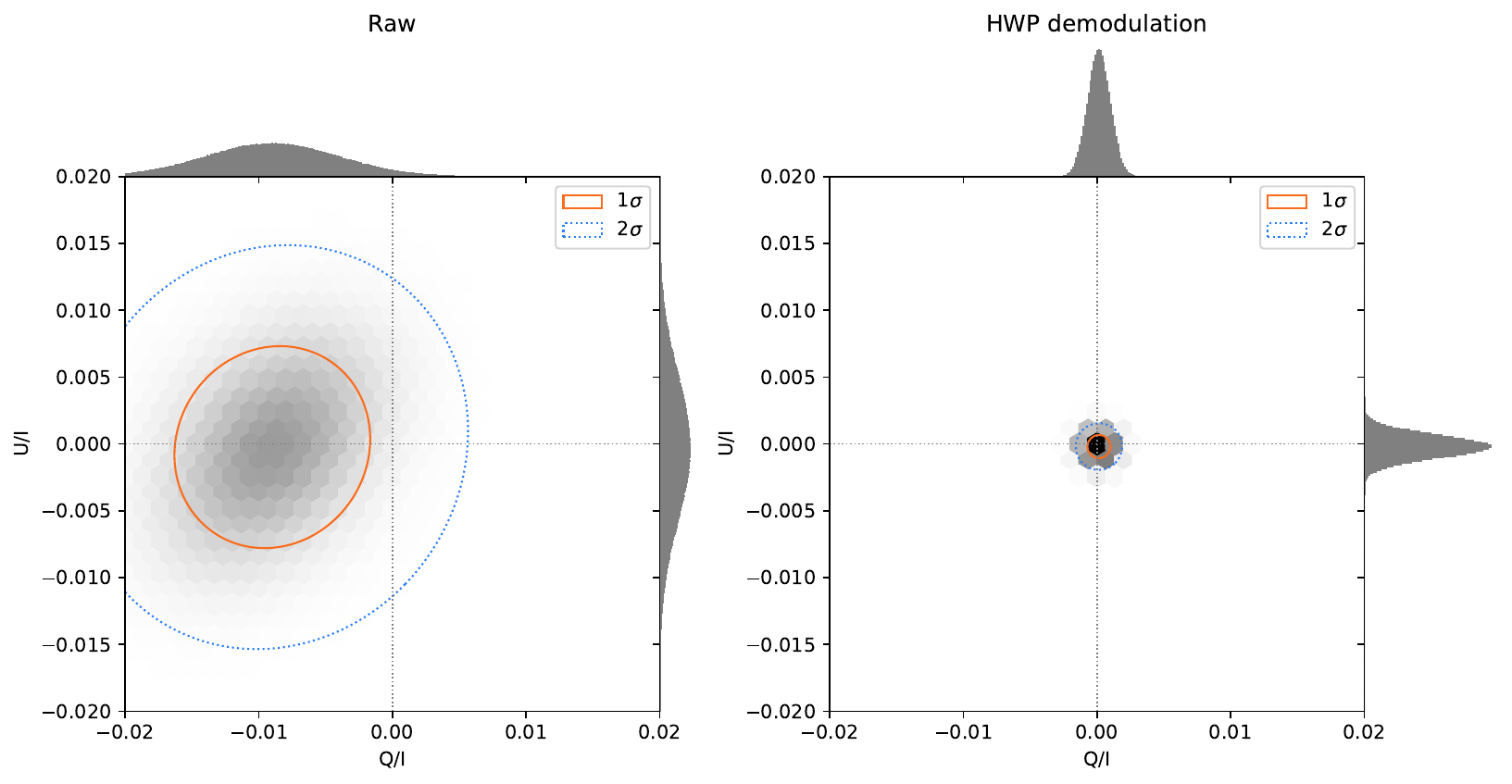}
	\caption{Density plots of the retrieved $q$ and $u$ for illumination with unpolarized light. The data is uncorrected. In the left figure we only applied spatial demodulation and in the right figure we used temporal and spatial demodulation with help of a rotating HWP.}
	\label{fig:hwp accuracy}
\end{figure}

To investigate the effects of a rotating HWP on the polarimetric accuracy, we exchanged the polarizer in the collimated beam set-up as shown in figure \ref{fig:collimated beam set-up} with a HWP (Thorlabs WPH10M-445 1"), and inserted the FBH450-40 narrowband filter, a $40$nm narrowband filter. For this test, the Thorlabs Kiralux camera was used. The half-wave plate was rotated in steps of $22.5^\circ$ over a $360^\circ$ rotation, and at each half-wave plate angle, 50 frames were taken at gain $0$ and $2$s exposure time.
Dark measurements were done at the same temperature for the same number of frames and the same exposure time. A mean dark frame is subtracted before demodulation.\\
In figure \ref{fig:hwp accuracy}, results are shown for demodulation with four half-wave plate angles (right-hand side), as compared to super-pixel demodulation at one half-wave plate angle only (left-hand side). To create the heatmaps, all pixels on the sensor are used, and $1\sigma$ and $2\sigma$ intervals are shown. The mean value of $\sqrt{q^2+u^2}$ of a pixel is $0.0114$ for uncorrected data, and $0.0011$ for the data that is demodulated with the half-wave plate, which is a factor $10$ improvement on the polarimetric accuracy. The size of the confidence ellipse decreases by a factor $10$ as well. This showcases an increased polarimetric accuracy for all pixels, not just their average value. The more circular shape is also a result of improved accuracy.

\section{Comparison of the Polarimetric Sensitivity and Accuracy between using a HWP and Sensor Calibration}
\begin{figure}[b]
    \centering
    \includegraphics[width=\textwidth]{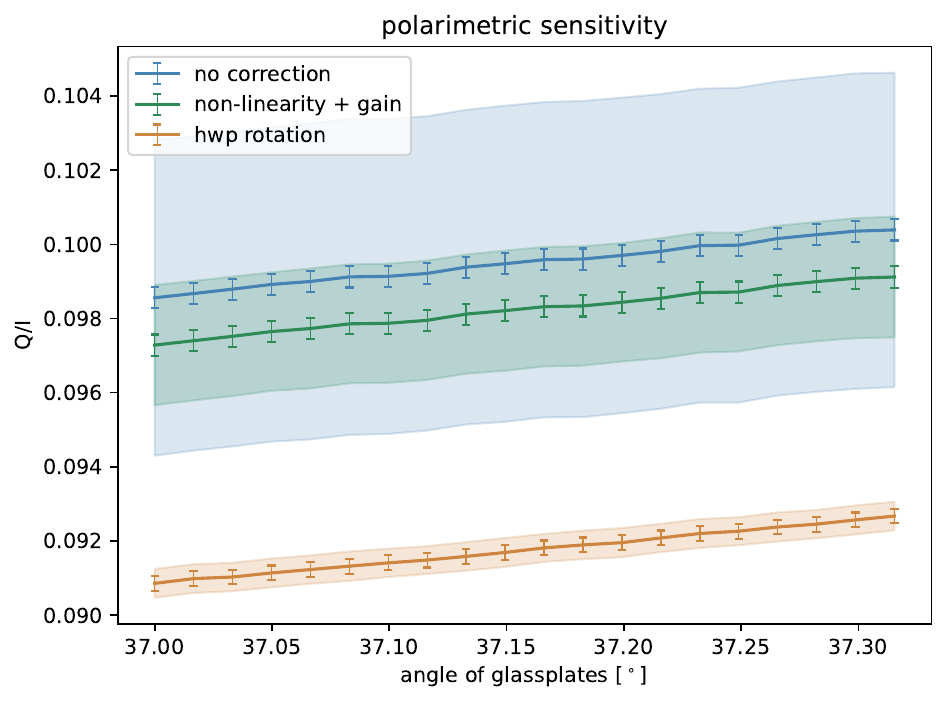}
    \caption{Polarimetric sensitivity measurements with the glassplates set-up, comparing uncorrected data (blue) to data corrected for pixel gain differences and linearity (green) and data that is averaged with a rotating HWP (orange). The averaging with a HWP shows an improvement in sensitivity with a factor $\sim10$. Error bars show the propagated standard error, while the shaded areas show the standard deviation in demodulated $Q/I$.}
  \label{fig:HWP sensitivity}
\end{figure}

To investigate the improvement in polarimetric sensitivity, the glassplate set-up was used that was described in section \ref{sec: glass plate}. One set of measurements was done with only the Thorlabs FBH550-40 narrowband filter (see figure \ref{fig:Glass_plates} for its location in the set-up), and for one set up images it was placed in a high precision motorized Thorlabs KPRM1E/M
rotating stage together with a Thorlabs achromatic half-wave plate AHWP10M-600, with operating range 400-800nm.
The monochrome FLIR was used for the measurement with gain $0$, and an exposure time of $250$ms. For the glassplates, 20 tilt angles were used from $37^\circ$ to $37.3154^\circ$ in steps of $0.0166^\circ$, to create polarization increments on the order of $10^{-4}$, so that sensitivity on this level can be tested for. To decrease the exected shot-noise to a similar level, 500 frames were taken without the half-wave plate, and 300 frames with the half-wave plate, at each of four rotational angles: $\beta\in(0^\circ,22.5^\circ,67.5^\circ,135^\circ)$. 500 dark frames were taken with the same settings, and the mean dark was subtracted from all science frames before demodulation.
Figure \ref{fig:HWP sensitivity} shows the results of the measurements with the glass plate set-up. 
In blue, the uncorrected data are shown, with simply $Q/I=\frac{I_0-I_{90}}{I_0+I_{90}}$. 
The error bars show the propagated standard error from the original frames and indicate the minimum expected random error on the data. 
The broader transparent blue band indicates the standard deviation found in the calculated $Q/I$-values, in a central region of 100x100 super-pixels on the detector. 
The fact that this range is so much wider than the expected random error, showcases the effect that all the different pixel properties have on the signal. 
It means that if one superpixel detects a signal of $Q/I\sim0.098$, its neighboring pixel could show a signal off by $~5\cdot10^{-3}$. 
Furthermore, even though the mean value is neatly defined, its accuracy is questionable. 
The orange plot, shown with the same errors, comes from the demodulation algorithm that includes the rotating half-wave plate as introduced in the theoretical part of this section. 
Two obvious changes occur: a linear translation to a slightly lower $Q/I$, but most importantly, the spread decreases by a factor $\sim10$ from an average of $4.2\cdot10^{-3}$ over all tilt angles to an average of $3.5\cdot10^{-4}$. This shows a significant improvement in polarimetric sensitivity. The green graph shows the same data as in blue, but corrected for both transmission and non-linearity based on the characterization of section \ref{sec:linearity}. This improves the polarimetric sensitivity from $4.2\cdot10^{-3}$ to $1.6\cdot10^{-3}$.

The difference between the partial correction and the results retrieved with the HWP indicates that further calibration beyond differential gain differences and non-linearity is still necessary. 

\section{Conclusion}

%

In this work we have performed a full calibration of four different MPA cameras with similar detector chips. 
We have shown through simulation and observation the effect of multiple effects that impact either the polarimetric accuracy, polarimetric sensitivity or both. 

Overall, performance over different cameras was more similar than the intrisic deviations over the arrays. 
Mainly, the accessibility of different features from the read-out electronics or specifications determined better polarimetric performance. 
The Thorlabs camera had reduced polarimetric accuracy due to the unpredictable dark behavior, and the FLIR cameras did have a larger dynamic range in possible exposure times, increasing their capabilities for different illumination strengths.

We found that different effects have different impacts on the polarimetric capabilities. 
We can summarize these effects as follows: 
\begin{itemize}
    \item \textbf{Uniform dark offset} induces a negative linear skewing as function of the degree of polarization of the order of $\sim$$10^{-3}$. This can be easily corrected for if taking the device temperature into account.
    \item \textbf{Spatial dark offset} is at most at the $\sim$$10^{-4}$ level, but can be easily corrected for by correctly estimating a framewide dark offset at the same temperaure.
    \item \textbf{Pixel gain differences} is of the order of $\sim$$10^{-2}$ level, with an immediate corresponding impact on the polarimetric zero point, and is therefore very important to correct by characterizing the effect per pixel under constant uniform unpolarized illumination. 
    \item \textbf{Non-Linearity} is only discussed at low illumination levels, and will skew polarization signals at very low digital signal levels. These are easily mitigated by boosting the gain in these regimes and consequently retrieving higher digital signals
    \item \textbf{MP angle misalignment} can have an impact of the order of $\sim$$10^{-2}$ for highly polarized sources. And can be corrected by characterizing the angles for all pixels. The actual algorithm of calibration depends on your demosaicing algorithm.
    \item \textbf{Extinction ratio} is also mainly important for highly polarized sources and can negatively skew your retrieved signal on the order of $\sim$$10^{-2}$ if not characterized per pixel. Again, the calibration algorithm depends on the demosaicing algorithm.
\end{itemize}
 
Next to the calibration of the MPA detectors by taking the effects above into account, we have performed a verification of a hyrbid modulation approach with and additional rotating half-wave plate. 
By changing the orientation of the half-wave plate, one can effectively exchange the orientation of the micro polarizers. 
The linear polarimetric signal coming from this temporal modulation, is theoretically insensitive to systematic pixel-specific effects. 
We found a difference between the linear polarimetric signal from partially calibrated measurements and measurements made using this temporal demodulation of the order of $5*10^{-3}$ which are significant seeing the corresponding sensitivity of the measurements. Which indicates that a full calibration is necessary. 

We conclude that while sub-percent accuracy and sensitivity is possible using the commercial MPA cameras with the SONY chips, calibration or additional modulation is vital to reduce a range of error sources to well below 1\% in terms of fractional linear polarization.
This conclusion opens up the use of the cameras for multiple interesting applications where high polarimetric accuracy is important. 
These applications include cloud/aerosol characterization, auroral measurements, and (indirectly) trace-gas monitoring.

\bibliographystyle{spiebib}
\bibliography{bibliography}

\begin{thebibliography}{10}

\bibitem{tyo_review_2006}
J.~S. Tyo, D.~L. Goldstein, D.~B. Chenault, and J.~A. Shaw, ``Review of passive imaging polarimetry for remote sensing applications,'' {\em Appl. Opt.}~{\bf 45}, p.~5453, 08 2006.

\bibitem{chenault_overview_2014}
F.~Snik, J.~Craven-Jones, M.~Escuti, S.~Fineschi, D.~Harrington, A.~De~Martino, D.~Mawet, J.~Riedi, and J.~S. Tyo, ``An overview of polarimetric sensing techniques and technology with applications to different research fields,'' p.~90990B, 05 2014.

\bibitem{rodenhuis_five-dimensional_2014}
M.~Rodenhuis, F.~Snik, G.~v. Harten, J.~Hoeijmakers, and C.~U. Keller, ``Five-dimensional optical instrumentation: combining polarimetry with time-resolved integral-field spectroscopy,'' in {\em Polarization: Measurement, Analysis, and Remote Sensing {XI}},   {\bf 9099}, pp.~158--173, {SPIE}, 06 2014.

\bibitem{snik_astronomical_2013}
F.~Snik and C.~U. Keller, ``Astronomical polarimetry: Polarized views of stars and planets,'' {\em Planets, Stars and Stellar Systems Volume 2: Astronomical Techniques, Software, and Data} , pp.~175--221, 2013.

\bibitem{bagnulo_stellar_2009}
S.~Bagnulo, M.~Landolfi, J.~D. Landstreet, E.~Landi~Degl'Innocenti, L.~Fossati, and M.~Sterzik, ``Stellar spectropolarimetry with retarder waveplate and beam splitter devices,'' {\em Publications of the Astronomical Society of the Pacific}~{\bf 121}, p.~993, 09 2009.
\newblock {ADS} Bibcode: 2009PASP..121..993B.

\bibitem{rust_integrated_1995}
D.~M. Rust, ``Integrated dual imaging detector,'' 08 1995.

\bibitem{kalayjian_polarization_1996}
Z.~Kalayjian, A.~G. Andreou, L.~Wolff, and N.~Sheppard, ``A polarization contrast retina using patterned iodine-doped {PVA} film,'' in {\em {ESSCIRC} '96: Proceedings of the 22nd European Solid-State Circuits Conference},  pp.~308--311, 09 1996.

\bibitem{guo_fabrication_2000}
J.~Guo and D.~Brady, ``Fabrication of thin-film micropolarizer arrays for visible imaging polarimetry,'' {\em Appl Opt}~{\bf 39}, pp.~1486--1492, 04 2000.

\bibitem{maruyama_32-mp_2018}
Y.~Maruyama, T.~Terada, T.~Yamazaki, Y.~Uesaka, M.~Nakamura, Y.~Matoba, K.~Komori, Y.~Ohba, S.~Arakawa, Y.~Hirasawa, Y.~Kondo, J.~Murayama, K.~Akiyama, Y.~Oike, S.~Sato, and T.~Ezaki, ``3.2-{MP} back-illuminated polarization image sensor with four-directional air-gap wire grid and 2.5- \${\textbackslash}mu\$ m pixels,'' {\em {IEEE} Trans. Electron Devices}~{\bf 65}, pp.~2544--2551, 06 2018.

\bibitem{momeni_analog_2006}
M.~Momeni and A.~Titus, ``An analog {VLSI} chip emulating polarization vision of octopus retina,'' {\em {IEEE} Trans. Neural Netw.}~{\bf 17}, pp.~222--232, 01 2006.

\bibitem{zhao_thin_2009}
X.~Zhao, F.~Boussaid, A.~Bermak, and V.~G. Chigrinov, ``Thin photo-patterned micropolarizer array for {CMOS} image sensors,'' {\em {IEEE} Photonics Technology Letters}~{\bf 21}, pp.~805--807, 06 2009.
\newblock Conference Name: {IEEE} Photonics Technology Letters.

\bibitem{gruev_ccd_2010}
V.~Gruev, R.~Perkins, and T.~York, ``{CCD} polarization imaging sensor with aluminum nanowire optical filters,'' {\em Opt. Express, {OE}}~{\bf 18}, pp.~19087--19094, 08 2010.
\newblock Publisher: Optica Publishing Group.

\bibitem{gruev_image_2008}
V.~Gruev, J.~Van~der Spiegel, and N.~Engheta, ``Image sensor with focal plane polarization sensitivity,'' in {\em 2008 {IEEE} International Symposium on Circuits and Systems ({ISCAS})},  pp.~1028--1031, 05 2008.
\newblock {ISSN}: 2158-1525.

\bibitem{chenault_infrared_2016}
D.~B. Chenault, J.~P. Vaden, D.~A. Mitchell, and E.~D. {DeMicco}, ``Infrared polarimetric sensing of oil on water,'' in {\em Remote Sensing of the Ocean, Sea Ice, Coastal Waters, and Large Water Regions 2016},   {\bf 9999}, pp.~89--100, {SPIE}, 10 2016.

\bibitem{zou_ultra-broadband_2020}
M.~Zou, M.~Su, and H.~Yu, ``Ultra-broadband and wide-angle terahertz polarization converter based on symmetrical anchor-shaped metamaterial,'' {\em Optical Materials}~{\bf 107}, p.~110062, 09 2020.

\bibitem{zhang_deep-ultraviolet_2019}
J.~Zhang, K.~Xie, Y.~Wei, P.~Wang, Z.~Hu, H.~Wang, F.~Wang, T.~Liang, S.~Ghafoor, H.~Jiang, L.~Zhang, and L.~Zhang, ``Deep-ultraviolet to mid-infrared polarizers by al nanowire metamaterials,'' {\em J. Phys. D: Appl. Phys.}~{\bf 52}, p.~365102, 07 2019.
\newblock Publisher: {IOP} Publishing.

\bibitem{hsu_full-stokes_2014}
W.-L. Hsu, G.~Myhre, K.~Balakrishnan, N.~Brock, M.~Ibn-Elhaj, and S.~Pau, ``Full-stokes imaging polarimeter using an array of elliptical polarizer,'' {\em Optics Express}~{\bf 22}, pp.~3063--3063, 02 2014.
\newblock Publisher: The Optical Society.

\bibitem{tu_division_2020}
X.~Tu, S.~{McEldowney}, Y.~Zou, M.~Smith, C.~Guido, N.~Brock, S.~Miller, L.~Jiang, and S.~Pau, ``Division of focal plane red-green-blue full-stokes imaging polarimeter,'' {\em Applied optics}~{\bf 59}, pp.~G33--G40, 08 2020.
\newblock Publisher: {NLM} (Medline).

\bibitem{guan_integrated_2022}
C.~Guan, R.~Zhang, J.~Chu, Z.~Liu, Y.~Fan, J.~Liu, and Z.~Yi, ``Integrated real-time polarization image sensor based on {UV}-{NIL} and calibration method,'' {\em {IEEE} Sensors Journal}~{\bf 22}, pp.~3157--3163, 02 2022.
\newblock Conference Name: {IEEE} Sensors Journal.

\bibitem{brock_snap-shot_2014}
N.~J. Brock, C.~Crandall, and J.~E. Millerd, ``Snap-shot imaging polarimeter: performance and applications,'' ~{\bf 9099}, p.~909903, 05 2014.
\newblock Conference Name: Polarization: Measurement, Analysis, and Remote Sensing {XI} {ADS} Bibcode: 2014SPIE.9099E..03B.

\bibitem{vorobiev_astronomical_2018}
D.~V. Vorobiev, Z.~Ninkov, and N.~Brock, ``Astronomical polarimetry with the {RIT} polarization imaging camera,'' {\em Publications of the Astronomical Society of the Pacific}~{\bf 130}(988), pp.~64501--64501, 2018.
\newblock Publisher: {IOP} Publishing.

\bibitem{millerd_vibration_2017}
J.~Millerd, N.~Brock, J.~Hayes, B.~Kimbrough, M.~North-Morris, and J.~C. Wyant, ``Vibration insensitive interferometry,'' ~{\bf 10567}, p.~105671P, 11 2017.
\newblock Conference Name: Society of Photo-Optical Instrumentation Engineers ({SPIE}) Conference Series {ADS} Bibcode: 2017SPIE10567E..1PM.

\bibitem{sun_unmanned_2024}
H.~Sun, G.~Sui, X.~Gu, Q.~Fu, H.~Shi, J.~Zhan, S.~Zhang, Y.~Li, and H.~Jiang, ``An unmanned aerial vehicle ({UAV})-borne dual-band polarization imaging system for evidence search,'' {\em Optics \& Laser Technology}~{\bf 169}, p.~109986, 02 2024.

\bibitem{takacs_polarized_2024}
P.~Takács, D.~Száz, B.~Bernáth, I.~Pomozi, and G.~Horváth, ``Polarized light pollution of fixed-tilt photovoltaic solar panels measured by drone-polarimetry and its visual-ecological importance,'' {\em Remote Sensing}~{\bf 16}, p.~1177, 03 2024.

\bibitem{wang_automatic_2023}
M.~Wang, S.~Qiu, W.~Jin, and J.~Yang, ``Automatic suppression method for water surface glints using a division of focal plane visible polarimeter,'' {\em Sensors (Basel)}~{\bf 23}, p.~7446, 08 2023.

\bibitem{inoue_motion-picture_2022}
T.~Inoue, K.~Nagao, A.~Matsunaka, T.~Kokubu, K.~Nishio, T.~Kubota, and Y.~Awatsuji, ``Motion-picture recording of light pulses with ultrashort time difference by polarization camera,'' {\em {IEEE} Photonics Technology Letters}~{\bf 34}, pp.~931--934, 09 2022.
\newblock Conference Name: {IEEE} Photonics Technology Letters.

\bibitem{huang_high-efficiency_2023}
F.~Huang, R.~Cao, P.~Lin, B.~Zhou, and X.~Wu, ``High-efficiency multispectral-polarization imaging system using polarization camera array with notch filters,'' {\em {IEEE} Trans. Instrum. Meas.}~{\bf 72}, pp.~1--13, 2023.

\bibitem{zhang_dual-dispersive_2024}
Y.~Zhang, H.~Li, J.~Sun, X.~Zhang, and Z.~Ling, ``Dual-dispersive spectral linear polarization imager based on coded-aperture,'' {\em Optics \& Laser Technology}~{\bf 170}, p.~110149, 03 2024.

\bibitem{tyo_total_2009}
J.~S. Tyo, C.~F. {LaCasse}, and B.~M. Ratliff, ``Total elimination of sampling errors in polarization imagery obtained with integrated microgrid polarimeters,'' {\em Opt. Lett.}~{\bf 34}, p.~3187, 10 2009.

\bibitem{ratliff_interpolation_2009}
B.~M. Ratliff, C.~F. {LaCasse}, and J.~S. Tyo, ``Interpolation strategies for reducing {IFOV} artifacts in microgrid polarimeter imagery,'' {\em Opt. Express, {OE}}~{\bf 17}, pp.~9112--9125, 05 2009.
\newblock Publisher: Optica Publishing Group.

\bibitem{ratliff_image_2006}
B.~M. Ratliff, J.~K. Boger, M.~P. Fetrow, J.~S. Tyo, and W.~T. Black, ``Image processing methods to compensate for {IFOV} errors in microgrid imaging polarimeters,'' in {\em Polarization: Measurement, Analysis, and Remote Sensing {VII}},   {\bf 6240}, pp.~106--117, {SPIE}, 05 2006.

\bibitem{hardie_super-resolution_2011}
R.~C. Hardie, D.~A. {LeMaster}, and B.~M. Ratliff, ``Super-resolution for imagery from integrated microgrid polarimeters,'' {\em Opt. Express, {OE}}~{\bf 19}, pp.~12937--12960, 07 2011.
\newblock Publisher: Optica Publishing Group.

\bibitem{mihoubi_survey_2018}
S.~Mihoubi, P.-J. Lapray, and L.~Bigué, ``Survey of demosaicking methods for polarization filter array images,'' {\em Sensors}~{\bf 18}, p.~3688, 11 2018.
\newblock Number: 11 Publisher: Multidisciplinary Digital Publishing Institute.

\bibitem{zhang_novel_2017}
J.~Zhang, W.~Ye, A.~Ahmed, Z.~Qiu, Y.~Cao, and X.~Zhao, ``A novel smoothness-based interpolation algorithm for division of focal plane polarimeters,'' in {\em 2017 {IEEE} International Symposium on Circuits and Systems ({ISCAS})},  pp.~1--4, 05 2017.
\newblock {ISSN}: 2379-447X.

\bibitem{gilboa_image_2014}
E.~Gilboa, J.~P. Cunningham, A.~Nehorai, and V.~Gruev, ``Image interpolation and denoising for division of focal plane sensors using gaussian processes,'' {\em Opt Express}~{\bf 22}, pp.~15277--15291, 06 2014.

\bibitem{wang_image_2018}
B.~Wang, S.~Li, W.~Ye, A.~Abubakar, X.~Pan, and X.~Zhao, ``Image denoising algorithms for {DoFP} polarization image sensors with non-gaussian noises,'' in {\em 2018 {IEEE} Asia Pacific Conference on Circuits and Systems ({APCCAS})},  pp.~419--422, 10 2018.

\bibitem{bai_noise_2022}
C.~Bai, Z.~Jiang, J.~Zhao, S.~Wu, and Q.~Zhang, ``Noise analysis in stokes parameter reconstruction for division-of-focal-plane polarimeters,'' {\em Appl. Opt., {AO}}~{\bf 61}, pp.~7084--7094, 08 2022.
\newblock Publisher: Optica Publishing Group.

\bibitem{liu_new_2020}
S.~Liu, J.~Chen, Y.~Xun, X.~Zhao, and C.-H. Chang, ``A new polarization image demosaicking algorithm by exploiting inter-channel correlations with guided filtering,'' {\em {IEEE} Trans. on Image Process.}~{\bf 29}, pp.~7076--7089, 2020.

\bibitem{morimatsu_monochrome_2020}
M.~Morimatsu, Y.~Monno, M.~Tanaka, and M.~Okutomi, ``Monochrome and color polarization demosaicking using edge-aware residual interpolation,'' {\em Proceedings - International Conference on Image Processing, {ICIP}}~{\bf 2020-Octob}, pp.~2571--2575, 2020.

\bibitem{morimatsu_monochrome_2021}
M.~Morimatsu, Y.~Monno, M.~Tanaka, and M.~Okutomi, ``Monochrome and color polarization demosaicking based on intensity-guided residual interpolation,'' {\em {IEEE} Sensors Journal}~{\bf 21}, pp.~26985--26996, 12 2021.
\newblock Conference Name: {IEEE} Sensors Journal.

\bibitem{liu_modified_2022}
X.~Liu, L.~Yang, and L.~Wang, ``Modified newton-residual interpolation for division of focal plane polarization image demosaicking,'' {\em Opt. Express, {OE}}~{\bf 30}, pp.~33048--33067, 08 2022.
\newblock Publisher: Optica Publishing Group.

\bibitem{yang_residual_2022}
J.~Yang, W.~Jin, S.~Qiu, F.~Xue, and M.~Wang, ``Residual interpolation integrated pixel-by-pixel adaptive iterative process for division of focal plane polarimeters,'' {\em Sensors}~{\bf 22}, p.~1529, 01 2022.
\newblock Number: 4 Publisher: Multidisciplinary Digital Publishing Institute.

\bibitem{hagen_fourier-domain_2024}
N.~Hagen, T.~Stockmans, Y.~Otani, and P.~Buranasiri, ``Fourier-domain filtering analysis for color-polarization camera demosaicking,'' {\em Applied Optics}~{\bf 63}, p.~2314, 03 2024.
\newblock {ADS} Bibcode: 2024ApOpt..63.2314H.

\bibitem{lemaster_improved_2014}
D.~A. {LeMaster} and K.~Hirakawa, ``Improved microgrid arrangement for integrated imaging polarimeters,'' {\em Opt. Lett., {OL}}~{\bf 39}, pp.~1811--1814, 04 2014.
\newblock Publisher: Optica Publishing Group.

\bibitem{alenin_optimal_2017}
A.~S. Alenin, I.~J. Vaughn, and J.~S. Tyo, ``Optimal bandwidth micropolarizer arrays,'' {\em Opt. Lett., {OL}}~{\bf 42}, pp.~458--461, 02 2017.
\newblock Publisher: Optica Publishing Group.

\bibitem{hao_new_2021}
J.~Hao, Y.~Wang, K.~Zhou, X.~Yu, and Y.~Yu, ``New diagonal micropolarizer arrays designed by an improved model in fourier domain,'' {\em Sci Rep}~{\bf 11}, p.~5778, 03 2021.

\bibitem{ratliff_alternative_2021}
B.~M. Ratliff and G.~C. Sargent, ``Alternative linear microgrid polarimeters: design, analysis, and demosaicing considerations,'' {\em Appl. Opt., {AO}}~{\bf 60}, pp.~5805--5818, 07 2021.
\newblock Publisher: Optica Publishing Group.

\bibitem{vaughn_spatio-temporal_2019}
I.~J. Vaughn and J.~S. Tyo, ``Spatio-temporal hybrid color-polarization channeled sensors,'' ~{\bf 11132}, p.~111320K, 09 2019.
\newblock Conference Name: Polarization Science and Remote Sensing {IX} {ADS} Bibcode: 2019SPIE11132E..0KV.

\bibitem{snik_multi-domain_2015}
F.~Snik, G.~van Harten, A.~S. Alenin, I.~J. Vaughn, and J.~S. Tyo, ``A multi-domain full-stokes polarization modulator that is efficient for 300-2500nm spectropolarimetry,'' ~{\bf 9613}, p.~96130G, 09 2015.
\newblock Conference Name: Polarization Science and Remote Sensing {VII} {ADS} Bibcode: 2015SPIE.9613E..0GS.

\bibitem{vaughn_portable_2015}
I.~J. Vaughn, O.~G. Rodríguez-Herrera, M.~Xu, and J.~S. Tyo, ``A portable imaging mueller matrix polarimeter based on a spatio-temporal modulation approach: theory and implementation,'' ~{\bf 9613}, p.~961312, 09 2015.
\newblock Conference Name: Polarization Science and Remote Sensing {VII} {ADS} Bibcode: 2015SPIE.9613E..12V.

\bibitem{gimenez-henriquez_characterization_2022}
Y.~C. Giménez-Henríquez, {\em Characterization and calibration of polarizer filter array stokes imaging systems}.
\newblock phdthesis, 03 2022.

\bibitem{york_characterization_2012}
T.~York and V.~Gruev, ``Characterization of a visible spectrum division-of-focal-plane polarimeter,'' {\em Appl Opt}~{\bf 51}, pp.~5392--5400, 08 2012.

\bibitem{powell_calibration_2013}
S.~B. Powell and V.~Gruev, ``Calibration methods for division-of-focal-plane polarimeters,'' {\em Opt Express}~{\bf 21}, pp.~21039--21055, 09 2013.

\bibitem{gimenez_calibration_2019}
Y.~Gimenez, P.-J. Lapray, A.~Foulonneau, and L.~Biguea, ``Calibration for polarization filter array cameras: recent advances,'' in {\em Fourteenth International Conference on Quality Control by Artificial Vision},  C.~Cudel, S.~Bazeille, and N.~Verrier, eds., p.~79, {SPIE}, 07 2019.

\bibitem{gimenez_calibration_2020}
Y.~Gimenez, P.-J. Lapray, A.~Foulonneau, and L.~Bigué, ``Calibration algorithms for polarization filter array camera: survey and evaluation,'' {\em {JEI}}~{\bf 29}, p.~041011, 03 2020.
\newblock Publisher: {SPIE}.

\bibitem{lane_calibration_2022}
C.~Lane, D.~Rode, and T.~Rösgen, ``Calibration of a polarization image sensor and investigation of influencing factors,'' {\em Appl. Opt., {AO}}~{\bf 61}, pp.~C37--C45, 02 2022.
\newblock Publisher: Optica Publishing Group.

\bibitem{hagen_calibration_2019}
N.~A. Hagen, S.~Shibata, and Y.~Otani, ``Calibration and performance assessment of microgrid polarization cameras,'' {\em {OE}}~{\bf 58}, p.~082408, 02 2019.
\newblock Publisher: {SPIE}.

\bibitem{hagen_generating_2019}
N.~Hagen, S.~Shibata, and Y.~Otani, ``Generating high-performance polarization measurements with low-performance polarizers: demonstration with a microgrid polarization camera,'' {\em {OE}}~{\bf 58}, p.~080501, 08 2019.
\newblock Publisher: {SPIE}.

\bibitem{rodriguez_practical_2022}
J.~Rodriguez, L.~Lew-Yan-Voon, R.~Martins, and O.~Morel, ``A practical calibration method for {RGB} micro-grid polarimetric cameras,'' {\em {IEEE} Robot. Autom. Lett.}~{\bf 7}, pp.~9921--9928, 10 2022.

\bibitem{szaz_drone-based_2023}
D.~Száz, P.~Takács, B.~Bernáth, G.~Kriska, A.~Barta, I.~Pomozi, and G.~Horváth, ``Drone-based imaging polarimetry of dark lake patches from the viewpoint of flying polarotactic insects with ecological implication,'' {\em Remote Sensing}~{\bf 15}, p.~2797, 05 2023.

\bibitem{venkatesulu_measuring_2022}
E.~Venkatesulu and J.~A. Shaw, ``Measuring the spectral response of a division-of-focal-plane polarization imager using a grating monochromator,'' {\em Appl Opt}~{\bf 61}, pp.~2364--2370, 03 2022.

\bibitem{roussel_polarimetric_2018}
S.~Roussel, M.~Boffety, and F.~Goudail, ``Polarimetric precision of micropolarizer grid-based camera in the presence of additive and poisson shot noise,'' {\em Opt. Express, {OE}}~{\bf 26}, pp.~29968--29982, 11 2018.
\newblock Publisher: Optica Publishing Group.

\bibitem{chen_analysis_2021}
Y.~Chen, Z.~Zhu, Z.~Liang, L.~E. Iannucci, S.~P. Lake, and V.~Gruev, ``Analysis of signal-to-noise ratio of angle of polarization and degree of polarization,'' {\em {OSA} Continuum, {OSAC}}~{\bf 4}, pp.~1461--1472, 05 2021.
\newblock Publisher: Optica Publishing Group.

\bibitem{snik_spectral_2009}
F.~Snik, T.~Karalidi, and C.~U. Keller, ``Spectral modulation for full linear polarimetry,'' {\em Appl. Opt.}~{\bf 48}, pp.~1337--1346, 03 2009.
\newblock Publisher: {OSA}.

\bibitem{smit_spex_2019}
J.~M. Smit, J.~H.~H. Rietjens, G.~van Harten, A.~Di~Noia, W.~Laauwen, B.~E. Rheingans, D.~J. Diner, B.~Cairns, A.~Wasilewski, K.~D. Knobelspiesse, R.~Ferrare, and O.~P. Hasekamp, ``{SPEX} airborne spectropolarimeter calibration and performance,'' {\em Applied Optics}~{\bf 58}, p.~5695, 07 2019.
\newblock {ADS} Bibcode: 2019ApOpt..58.5695S.

\bibitem{rietjens_spexone_2023}
J.~Rietjens, M.~Smit, J.~Campo, T.~Bouchan, R.~Cooney, P.~Piron, P.~Tol, R.~Laasner, R.~van Hees, J.~Landgraf, and O.~Hasekamp, ``{SPEXone} multi-angle spectropolarimeter characterization, calibration, and key data derivation using the l0-1b processor,'' ~{\bf 12777}, p.~127774U, 07 2023.
\newblock Conference Name: Society of Photo-Optical Instrumentation Engineers ({SPIE}) Conference Series {ADS} Bibcode: 2023SPIE12777E..4UR.

\bibitem{hasekamp_retrieval_2007}
O.~P. Hasekamp and J.~Landgraf, ``Retrieval of aerosol properties over land surfaces: capabilities of multiple-viewing-angle intensity and polarization measurements,'' {\em Applied Optics}~{\bf 46}, pp.~3332--3344, 06 2007.
\newblock {ADS} Bibcode: 2007ApOpt..46.3332H.

\bibitem{dubovik_polarimetric_2019}
O.~Dubovik, Z.~Li, M.~I. Mishchenko, D.~Tanré, Y.~Karol, B.~Bojkov, B.~Cairns, D.~J. Diner, W.~R. Espinosa, P.~Goloub, X.~Gu, O.~Hasekamp, J.~Hong, W.~Hou, K.~D. Knobelspiesse, J.~Landgraf, L.~Li, P.~Litvinov, Y.~Liu, A.~Lopatin, T.~Marbach, H.~Maring, V.~Martins, Y.~Meijer, G.~Milinevsky, S.~Mukai, F.~Parol, Y.~Qiao, L.~Remer, J.~Rietjens, I.~Sano, P.~Stammes, S.~Stamnes, X.~Sun, P.~Tabary, L.~D. Travis, F.~Waquet, F.~Xu, C.~Yan, and D.~Yin, ``Polarimetric remote sensing of atmospheric aerosols: Instruments, methodologies, results, and perspectives,'' {\em Journal of Quantitative Spectroscopy and Radiative Transfer}~{\bf 224}, pp.~474--511, 2019.

\bibitem{mcbride_spatial_2019}
B.~{McBride}, J.~V. Martins, H.~Barbosa, W.~Birmingham, and L.~Remer, ``Spatial distribution of cloud droplet size properties from airborne hyper-angular rainbow polarimeter ({AirHARP}) measurements,'' {\em Atmospheric Measurement Techniques Discussions} , pp.~1--32, 2019.

\bibitem{levis_3d_2021}
A.~Levis, A.~B. Davis, J.~R. Loveridge, and Y.~Y. Schechner, ``3d cloud tomography and droplet size retrieval from multi-angle polarimetric imaging of scattered sunlight from above,'' ~{\bf 11833}, p.~1183305, 08 2021.
\newblock Conference Name: Polarization Science and Remote Sensing X {ADS} Bibcode: 2021SPIE11833E..05L.

\bibitem{lilensten_thermospheric_2013}
J.~Lilensten, M.~Barthélémy, P.-O. Amblard, H.~Lamy, C.~S. Wedlund, V.~Bommier, J.~Moen, H.~Rothkaehl, J.~Eymard, and J.~Ribot, ``The thermospheric auroral red line polarization: confirmation of detection and first quantitative analysis,'' {\em J. Space Weather Space Clim.}~{\bf 3}, p.~A01, 2013.
\newblock Publisher: {EDP} Sciences.

\bibitem{barthelemy_measurement_2019}
M.~Barthelemy, H.~Lamy, A.~Vialatte, M.~G. Johnsen, G.~Cessateur, and N.~Zaourar, ``Measurement of the polarisation in the auroral n2+ 427.8 nm band,'' {\em Journal of Space Weather and Space Climate}~{\bf 9}, p.~A26, 05 2019.
\newblock {ADS} Bibcode: 2019JSWSC...9A..26B.

\bibitem{bosse_nightglow_2020}
L.~Bosse, J.~Lilensten, N.~Gillet, S.~Rochat, A.~Delboulbe, S.~Curaba, A.~Roux, Y.~Magnard, M.~G. Johnsen, U.-P. Lovhaug, P.-O. Amblard, N.~Le~Bihan, M.~Nabon, H.~Marif, F.~Auriol, and C.~Nous, ``On the nightglow polarisation for space weather exploration,'' {\em J. Space Weather Space Clim.}~{\bf 10}, p.~35, 07 2020.
\newblock Place: Les Ulis Cedex A Publisher: Edp Sciences S A {WOS}:000556678200003.

\bibitem{derkink_imaging_2022}
F.~M. Derkink, D.~Doelman, D.~van Dam, B.~J. Brenny, B.~Speet, H.~van Brug, C.~Keller, and F.~Snik, ``Imaging trace gas concentrations with a compact snapshot device that converts their spectral features into a polarization signal,'' ~{\bf 12235}, p.~122350L, 09 2022.
\newblock Conference Name: Imaging Spectrometry {XXV}: Applications, Sensors, and Processing {ADS} Bibcode: 2022SPIE12235E..0LD.

\bibitem{mcclain_depolarization_1995}
S.~C. {McClain}, C.~L. Bartlett, J.~L. Pezzaniti, and R.~A. Chipman, ``Depolarization measurements of an integrating sphere,'' {\em Appl. Opt., {AO}}~{\bf 34}, pp.~152--154, 01 1995.
\newblock Publisher: Optica Publishing Group.

\bibitem{chipman_polarization_1986}
R.~Chipman, ``Polarization aberrations of lenses,'' in {\em 1985 International Lens Design Conference},   {\bf 0554}, pp.~82--87, {SPIE}, 02 1986.

\bibitem{harten_calibration_2018}
G.~v. Harten, D.~J. Diner, B.~J.~S. Daugherty, B.~E. Rheingans, M.~A. Bull, F.~C. Seidel, R.~A. Chipman, B.~Cairns, A.~P. Wasilewski, K.~D. Knobelspiesse, G.~van Harten, D.~J. Diner, B.~J.~S. Daugherty, B.~E. Rheingans, M.~A. Bull, F.~C. Seidel, R.~A. Chipman, B.~Cairns, A.~P. Wasilewski, and K.~D. Knobelspiesse, ``Calibration and validation of airborne multiangle {SpectroPolarimetric} imager ({AirMSPI}) polarization measurements,'' {\em Applied Optics}~{\bf 57}, pp.~4499--4499, 06 2018.
\newblock Publisher: {OSA}.

\bibitem{wang_noise_2008}
X.~Wang, {\em Noise in sub-micron {CMOS} image sensors: Proefschrift}, 2008.

\bibitem{widenhorn_temperature_2002}
R.~Widenhorn, M.~M. Blouke, A.~Weber, A.~Rest, and E.~Bodegom, ``Temperature dependence of dark current in a {CCD},'' pp.~193--201, 04 2002.

\bibitem{li_optimal_2019}
X.~Li, H.~Hu, M.~Boffety, S.~Roussel, T.~Liu, and F.~Goudail, ``Optimal tradeoff between precision and sampling rate in {DoFP} imaging polarimeters,'' {\em Opt. Lett., {OL}}~{\bf 44}, pp.~5900--5903, 12 2019.
\newblock Publisher: Optica Publishing Group.

\bibitem{shibata_robust_2019}
S.~Shibata, N.~Hagen, and Y.~Otani, ``Robust full stokes imaging polarimeter with dynamic calibration,'' {\em Opt. Lett., {OL}}~{\bf 44}, pp.~891--894, 02 2019.
\newblock Publisher: Optica Publishing Group.

\bibitem{shibata_video-rate_2019}
S.~Shibata, M.~Suzuki, N.~Hagen, and Y.~Otani, ``Video-rate full-stokes imaging polarimeter using two polarization cameras,'' {\em {OE}}~{\bf 58}, p.~103103, 10 2019.
\newblock Publisher: {SPIE}.

\bibitem{li_vortex_2024}
X.~Li and F.~Goudail, ``Vortex retarder-based stokes polarimeters: optimal data processing and autocalibration capability,'' {\em Opt. Lett., {OL}}~{\bf 49}, pp.~1696--1699, 04 2024.
\newblock Publisher: Optica Publishing Group.

\end{thebibliography}
\end{document}